\documentclass[twocolumn,showpacs,pra,preprintnumbers,amsmath,amssymb,superscriptaddress]{revtex4}

\usepackage{graphicx}
\usepackage{dcolumn}
\usepackage{bm}
\usepackage{color}
\usepackage{braket}

\begin{document}


\title{Terahertz radiation induced by coherent phonon generation via impulsive stimulated Raman scattering in paratellurite}

\author{M. Sotome}
\affiliation{Department of Advanced Materials Science, The University of Tokyo, 5-1-5 Kashiwa-no-ha, Chiba 277-8561, Japan}

\author{N. Kida}
\affiliation{Department of Advanced Materials Science, The University of Tokyo, 5-1-5 Kashiwa-no-ha, Chiba 277-8561, Japan}

\author{R. Takeda}
\affiliation{Department of Advanced Materials Science, The University of Tokyo, 5-1-5 Kashiwa-no-ha, Chiba 277-8561, Japan}

\author{H. Okamoto}
\affiliation{Department of Advanced Materials Science, The University of Tokyo, 5-1-5 Kashiwa-no-ha, Chiba 277-8561, Japan}

\date{\today}

\begin{abstract} 
We report on the observation of terahertz radiation in a non-centrosymmetric insulating oxide, paratellurite ($\alpha$-TeO$_2$) by irradiation of a femtosecond laser pulse at room temperature. In the power spectrum of the terahertz radiation, an intensity fringe pattern with a period of $\sim$ 0.25 THz shows up below 3 THz. It can be reproduced by taking into account the effective generation length for the terahertz radiation with a poor phase-matching condition. In addition, a temporal oscillation component appears in the radiated terahertz wave with a frequency of $\sim$ 3.71 THz, which is in good agreement with the center frequency of the Raman active longitudinal optical (LO) $E$ mode. On the basis of comprehensive polarized optical and Raman spectroscopic studies, we explain the generation mechanism of the temporal oscillation component in terms of the coherent phonon generation via impulsive stimulated Raman scattering.
\end{abstract}



\maketitle

\section{Introduction}

An ultrafast laser pulse, whose pulse width is shorter than the inverse of the frequency of the optical phonon mode, makes it possible to coherently drive optical phonons \cite{MCardona,TDekorsyREV,RMerlin}. This process is termed as coherent phonon generation. Interesting phenomena based upon coherent phonon generation were previously reported by a variety of experiments \cite{TDekorsyREV,RMerlin}, for examples, ultrafast control of coherent phonons by using a pair of femtosecond laser pulses with an appropriate interval \cite{TDekorsy3,PCMPlanken}, the observation of vibronic wavepackets in time domain \cite{AGambetta}, and the generation of the frequency comb over 100 THz \cite{MHase}. Furthermore, it was demonstrated that narrowband terahertz waves can be emitted by coherent phonon generation in various kinds of semiconductors such as Te \cite{TDekorsy1,TDekorsy2}, GaAs \cite{AVKuznetsov}, GaAs/AlAs multiple quantum wells \cite{MNakayama}, (Pb,Cd)Te \cite{MTani}, InSb \cite{PGu,PGu2,MPHasselbeck}, and Li(In,Ga)(S,Se) \cite{KTakeya}. 

In these semiconductors, it was reported that the generated terahertz wave consists of a single-cycle terahertz pulse and a temporal oscillation component; the frequency of the later corresponds to that of the optical phonon mode. So far, two mechanisms of terahertz radiation by the coherent phonons have been proposed. One is the generation of the coherent longitudinal optical (LO) phonons of infrared active modes \cite{AVKuznetsov}; the generated optical phonons modulate nonlinear polarization $P_{\rm NL}$, resulting in the emission of the terahertz wave at the phonon frequency. In optically opaque semiconductors such as Te and GaAs, the modulation of $P_{\rm NL}$ is induced mainly by photogenerated carriers in the screening process \cite{TDekorsy1,TDekorsy2,MNakayama,MTani,PGu,PGu2}. Especially, in the case of Te, it was confirmed that a narrowband terahertz wave was due to the generation of the coherent phonon of the infrared active LO $A_1$ mode \cite{TDekorsy1}. The other proposed mechanism is the generation of the coherent LO phonons of the Raman active modes via impulsive stimulated Raman scattering (ISRS) \cite{YXYan}. In InSb \cite{MPHasselbeck} and Li(In,Ga)(S,Se) \cite{KTakeya}, ISRS was proposed to explain the observed oscillation components in the radiated terahertz waves. However, the role of the ISRS mechanism in terahertz radiation has not been clarified yet. In the optically opaque semiconductors such as InSb, $P_{\rm NL}$ would be modulated by the screening process of the photogenerated carriers. Thus, it is difficult to isolate the contribution of ISRS to the measured terahertz radiation. Furthermore, an infrared active mode usually appears at nearly the same frequency of the Raman active mode in non-centrosymmetric media. Thus, the comprehensive optical and Raman spectroscopic studies with high frequency resolution are necessary to identify the terahertz radiation mechanism as coherent phonon generation by ISRS.

$\alpha$-TeO$_2$ is an ideal compound to scrutinize the role of ISRS for the terahertz radiation. In contrast to the optically opaque semiconductors such as Te and InSb, $\alpha$-TeO$_2$ is an insulator with a transparent window from 0.33 $\mu$m to 6.5 $\mu$m \cite{JLiebertz,NUchida}. Thus, no photo-carriers are generated by the irradiation of femtosecond laser pulses with a wavelength of 800 nm used in this work. Crystal structure of $\alpha$-TeO$_2$ is tetragonal with the point group of $D_4$ (422) \cite{PAThomas}. Figure 1(a) shows the schematic illustration of the crystal structure \cite{RWGWyckoff}. $\alpha$-TeO$_2$ contains four TeO$_2$ molecules in the unit cell, as guided by the dotted lines. We also illustrate the crystal view of (110) and (101) planes in Figs. 1(b) and (c), respectively. At room temperature, $\alpha$-TeO$_2$ lacks the space inversion symmetry, resulting in the piezoelectricity \cite{GArlt}. Due to the large acoustic-optic effect \cite{GArlt,UchidaOhmachi}, $\alpha$-TeO$_2$ is used as a constitutive element for the optical modulation and deflection devices in the visible frequency region. Furthermore, it was reported that the second-order nonlinear optical coefficient of $\alpha$-TeO$_2$ has a large value in the transparent frequency region, e.g., 0.59 pm/V \cite{DSChemla} or 0.69 $\pm$ 0.08 pm/V at 1.064 $\mu$m \cite{SSingh}. However, terahertz radiation in $\alpha$-TeO$_2$ has never been reported so far.

In this paper, we report on the first observation of the terahertz radiation in $\alpha$-TeO$_2$ by the irradiation of a femtosecond laser pulse at room temperature. In the power spectrum of terahertz radiation, an intensity fringe pattern appeared below 3 THz with a period of $\sim$ 0.25 THz. In addition, we observed a temporal oscillation component with a frequency of $\sim$ 3.71 THz in the radiated terahertz wave. The frequency of this component just corresponds to the center frequency of a Raman active $E_{\rm LO}$ mode. On the basis of the systematic investigation of polarized optical and Raman spectroscopies in the terahertz frequency region, we concluded that the observed temporal oscillation component in the radiated terahertz wave is due to the coherent phonon generation via ISRS.

The format of this article is as follows. In Sec. II, we briefly describe the general expression of the second-order nonlinear optical susceptibility in $\alpha$-TeO$_2$ and examine the optical rectification for the terahertz radiation. We detail the experimental setups in Sec. III. Section IV is devoted to show the experimental results. After the detailed descriptions of the results of the polarized optical spectroscopy in the visible and terahertz frequency regions (Sec. IV. A) and the polarized Raman spectroscopy (Sec. IV. B), we provide the basic features of terahertz radiation in $\alpha$-TeO$_2$ in Sec. IV. C and show the results of its azimuth-angle and laser-power dependences in Sec. IV. D. In Sec. V, we discuss the radiation mechanism of the single-cycle terahertz pulse at 0 ps in terms of the optical rectification (Sec. V. A), the origins of the intensity fringe pattern in the power spectrum of terahertz radiation (Sec. V. B), the second pulse at around 3.5 ps (Sec. V. C), and the temporal oscillation component with a frequency of $\sim$ 3.71 THz (Sec. V. D). Summary is given in Sec. VI.

\section{Second-order nonlinear optical susceptibility tensor}

In this section, we briefly describe the general mechanism of the laser-induced terahertz radiation in non-centrosymmetric media and summarize the non-zero components of the second-order nonlinear optical susceptibility tensor in $\alpha$-TeO$_2$. When a non-centrosymmetric medium is irradiated with a femtosecond laser pulse, a transient nonlinear polarization $P_{\rm NL}$ is induced in the subpicosecond time scale. This results in the emission of terahertz waves into free space. Such an optical rectification has been widely recognized as the dominant process of terahertz radiation in various non-centrosymmetric media such as ZnTe \cite{MTonouchi_Rev}. In the case of $\alpha$-TeO$_2$, the non-zero components of the second-order nonlinear optical susceptibility tensor are $\chi^{(2)}_{yzx} = \chi^{(2)}_{yxz}= -\chi^{(2)}_{xyz}=-\chi^{(2)}_{xzy}$ \cite{SSingh,JFNye,YRShen}. Then, $P_{\rm NL}$ using the contradicted $d$ tensor is given by
\begin{eqnarray}
\begin{array}{ccc}
P_{\rm NL} & = & \epsilon_0
\left[
\begin{array}{cccccc}
0 & 0 & 0 & d_{14} & 0 &0 \\
0 & 0 & 0 & 0 & -d_{14} &0 \\
0 & 0 & 0 & 0 & 0 &0 \\
\end{array}
\right]
\left[
\begin{array}{c}
E_x^2\\
E_y^2\\
E_z^2\\
2E_yE_z\\
2E_zE_x\\
2E_xE_y\\
\end{array}
\right].
\end{array}
\end{eqnarray}
Here, $\epsilon_0$ is the vacuum permittivity. $E_x$, $E_y$, and $E_z$ represent $x$-, $y$-, and $z$-components of the electric field of a femtosecond laser pulse, respectively. We used the laboratory coordinate as $X$-(horizontal) axis and $Y$-(vertical) axis [see Fig. 5(a)]. The angle $\theta$ was defined as the angle of the [010] direction of the crystal relative to the $X$-axis. When the electric field $E_0$ of a femtosecond laser pulse is set parallel to the $X$-axis ($E_0\parallel X$), $P_{\rm NL}$ of a (101)-oriented single crystal as a function of $\theta$ is given by,
\begin{eqnarray}
P_{\rm NL}=
\begin{array}{ccl}
\left[ 
\begin{array}{c}
P_{1\overline{1}0}\\
P_{010}\\
P_{101}\\
\end{array} 
\right]
=
\epsilon _{0} d_{14} E^{2}_{0}
\left[ 
\begin{array}{c}
-\sqrt{2}\sin\theta\cos\theta\\
\sin^2\theta\\
0\\
\end{array} 
\right]. 
\end{array}
\end{eqnarray}
On the other hand, in the $E_0\parallel Y$ configuration, $P_{\rm NL}$ as a function of $\theta$ is given by,
\begin{eqnarray}
P_{\rm NL}=
\begin{array}{ccl}
\left[ 
\begin{array}{c}
P_{1\overline{1}0}\\
P_{010}\\
P_{101}\\
\end{array} 
\right]
=
\epsilon _{0} d_{14} E^{2}_{0}
\left[ 
\begin{array}{c}
\sqrt{2}\sin\theta\cos\theta\\
\cos^2\theta\\
0\\
\end{array} 
\right].
\end{array}
\end{eqnarray}

\section{Experimental Procedures}

We used commercially available 0.5-mm-thick (110)-oriented $\alpha$-TeO$_2$ single crystals to measure the optical spectra for extraordinary ray ($E^\omega\parallel [001]$) and ordinary ray ($E^\omega\parallel [1\overline{1}0]$), respectively. In the terahertz frequency region, we measured the reflectance and transmittance spectra by using Fourier-transformed infrared (FT-IR) spectroscopy and terahertz time-domain spectroscopy, respectively. The frequency range covered by FT-IR spectrometer was 40--7000 cm$^{-1}$. In the terahertz time-domain spectroscopy, we used a mode-locked Ti:Al$_2$O$_3$ laser (the center wavelength of 800 nm, the repetition rate of 80 MHz, and the pulse width of 100 fs). Terahertz emitter was a 0.5-mm thick (110)-oriented ZnTe single crystal. The terahertz wave passing through the sample was detected by a low-temperature-grown GaAs (LT-GaAs) detector. For the measurements of the reflectance and transmittance in the energy range of 0.5--5 eV, we used the grating monochromator equipped with an optical microscope. 

For the polarized Raman scattering experiments, commercially available 0.5-mm-thick (110)- and (101)-oriented $\alpha$-TeO$_2$ single crystals were used. The sample was excited by a continuous wave He-Ne laser [632 nm (1.96 eV)]. The scattering light was analyzed by a triple-grating monochromator and detected by a CCD camera in the backscattering geometry.

The terahertz radiation experiments were carried out on a (101)-oriented $\alpha$-TeO$_2$ single crystal in a dried air box to eliminate the absorption of the terahertz waves by water vapor. We adopted standard photoconducting sampling technique with a LT-GaAs detector \cite{MTonouchi_Rev}. A beam from the mode-locked Ti:Al$_2$O$_3$ laser was focused on the sample in normal incidence with a beam spot diameter of $\sim$ 25 $\mu$m. Unless otherwise stated, the laser power was fixed to be 30 mW ($\sim$ 150 $\mu$J/cm$^2$ per pulse). We directly measured the waveform of the terahertz electric field by monitoring the photocurrent of the LT-GaAs detector. A wire grid polarizer was inserted in front of the LT-GaAs detector. The additional details of experimental procedures will be described in Sec IV. C and Sec. IV. D.

All the experiments were performed at room temperature.

\section{Results}
\subsection{Polarized optical spectra}

First, we detail the polarized optical spectra in the terahertz, visible, and ultraviolet frequency regions in Fig. 2. Figure 2(e) shows the polarized reflectance $R$ spectra in 0.5--5 eV. $R$ spectra in $E^\omega\parallel [001]$ (blue line) and $E^\omega\parallel [1\overline{1}0]$ (red line) configurations were obtained using a (110)-oriented single crystal. We also show in Fig. 2(f) the polarized transmittance $T$ spectra in these configurations. The steep decrease in $T$ and the gradual increase in $R$ above 3.7 eV are due to the presence of the interband transition from O $2p$-state to Te $5s$-state at $\sim$ 4 eV \cite{NUchida,TTakizawa,JRobertson}. In $E^\omega\parallel [001]$ configuration, a peak structure at 4.28 eV is seen in $R$ spectrum. On the other hand, a peak structure at 4.81 eV with a shoulder structure at 4.26 eV is discerned in $E^\omega\parallel [1\overline{1}0]$ configuration. These structures were previously reported in Ref. \onlinecite{TTakizawa} and assigned to band-edge excitons. In order to obtain the refractive index spectra, we performed the dispersion analysis of the measured $R$ spectra; we assumed that $R$ spectrum is composed of Lorentz oscillators. The fitting results by assuming Lorentz oscillators with the damping rate $\gamma_i$ and the oscillator strength $f_i$ are shown by green broken lines in Fig. 2(e); the measured $R$ spectra were reproduced with parameters listed in Table I. With the use of the obtained fitting parameters, we calculated the refractive index spectra for ordinary ray $n_{\rm o}$ and extraordinary ray $n_{\rm e}$ in the visible to ultra-violet frequency regions, which are shown by red and blue lines in Fig. 2(g), respectively. There is refractive index anisotropy in the transparent frequency region below 3.6 eV; the birefringence is $\sim$ 0.16 at 800 nm. The obtained $n_{\rm o}$ and $n_{\rm e}$ spectra are consistent with those previously reported in the transparent frequency region (1.2--3.1 eV) \cite{NUchida}. We also evaluated the group refractive index $n_{\rm g}$ spectrum for $[10\overline{1}]$-polarized light with the use of the parameters extracted from the dispersion analysis of the $R$ spectrum. $n_{\rm g}$ is given by $| n-\lambda\frac{dn}{d\lambda}|$, where $\lambda$ is the wavelength and $n=(n_{\rm o}+n_{\rm e})/2$. The calculated $n_{\rm g}$ spectrum is shown by the black line in Fig. 2(g); $n_{\rm g}$ at 800 nm [the wavelength of the femtosecond laser used in the terahertz radiation experiments (the vertical broken line)] was estimated to be $\sim$ 2.52. As seen in the absorption coefficient $\alpha$ spectra [Fig. 2(h)], which were obtained by the relationship of $\alpha=-\frac{1}{d}\ln(\frac{T}{(1-R)^2})$ with the thickness $d$ of the sample (0.5 mm), $\alpha$-TeO$_2$ has a transparent window below 3.6 eV \cite{NUchida}, i.e., $\alpha$ at 800 nm for extraordinary and ordinary rays are negligibly small ($\sim$ 1 cm$^{-1}$).

Figure 2(b) shows the polarized $T$ spectra in $E^\omega\parallel [001]$ (circles) and $E^\omega\parallel [1\overline{1}0]$ (squares) configurations, measured by terahertz time-domain spectroscopy. Due to the presence of the strong absorptions, we only yielded the $T$ spectra below $\sim$ 2.0 THz in $E^\omega\parallel [001]$ configuration and below $\sim$ 3.0 THz in $E^\omega\parallel [1\overline{1}0]$ configuration. From the measured $T$ spectra, we derived $\alpha$ and $n$ spectra; the details of the derivation procedure are described in Appendix A. The $\alpha_{\rm o}$ and $\alpha_{\rm e}$ spectra are shown by squares and circles in Fig. 2(d), respectively. We also show in Fig. 2(c) the $n_{\rm o}$ and $n_{\rm e}$ spectra, which are represented by squares and circles, respectively. To obtain the optical spectra above 2.0 THz, we used the polarized $R$ spectra measured in $E^\omega\parallel [001]$ and $E^\omega\parallel [1\overline{1}0]$ configurations, which are shown by open circles and squares in Fig. 2(a), respectively. In $E^\omega\parallel [1\overline{1}0]$ configuration, a reflection peak is discerned at around 3.7 THz. On the other hand, a pronounced dispersion appears at around 3 THz in $E^\omega\parallel [001]$ configuration. We performed the dispersion analysis to extract complex optical constants. In this analysis, we used the polarized $R$ spectra in the frequency range of 1.9--24 THz [see Fig. 10(a)]; the analysis is detailed in Appendix B. The solid lines in Fig. 2(a) are the calculated $R$ spectra obtained by the dispersion analysis, which well reproduce the measured $R$ spectra. The $n$ and $\alpha$ spectra are also calculated and shown by solid lines in Figs. 2(c) and 2(d), respectively. The calculated $n$ and $\alpha$ (the broken lines) are in good agreement with those (symbols) obtained by the terahertz time-domain spectroscopy. This ensures the validity of the dispersion analysis used here. As can be seen, there is a strong anisotropy in the terahertz frequency region. A peak is discerned at 3.66 THz in the $\alpha_{\rm o}$ spectrum. Accordingly, a dispersive structure emerges in the $n_{\rm o}$ spectrum. On the other hand, $n_{\rm e}$ and $\alpha_{\rm e}$ show a more pronounced dispersion and a peak structure at 2.58 THz, respectively. The observed results are in good agreement with those previously reported in Ref. \onlinecite{DMKorn}.

According to the factor group analysis of $\alpha$-TeO$_2$, the representations of the optical phonons (excluding the acoustic phonons) at zone center are $4A_1+4A_2+5B_1+4B_2+8E$ \cite{DMKorn,ASPine}. We listed the polarization selection rules of the phonon modes in the terahertz frequency region in Table II. $A_2$ modes are Raman inactive and infrared active only for extraordinary ray. On the contrary, $E$ modes are Raman and infrared active for ordinary ray. Thus, the peak structures at 2.58 THz in the $\alpha_{\rm e}$ spectrum and 3.66 THz in the $\alpha_{\rm o}$ spectrum can be assigned to the $A_2$ and $E$ modes, respectively.

\subsection{Polarized Raman spectra}
So far, several results were reported on the polarized Raman spectra of $\alpha$-TeO$_2$ in the terahertz frequency region \cite{ASPine,MKrauzman,BAyrault,VRodriguez}. However, the mode assignment was controversial. For example, several forbidden Raman modes, which cannot be explained by the leakage of the polarization due to the misorientation of the crystal, show up when a (001)-oriented single crystal was used (Ref. \onlinecite{ASPine}). The `anomalous polarization-selection-rule violations' make it difficult to precisely perform the assignment of the observed Raman modes. Very recently, Raman and hyper-Raman scattering experiments using (110)-oriented single crystals were conducted to exclude the effect of the optical activity of $\alpha$-TeO$_2$ in the transparent frequency region \cite{VRodriguez}. The revised interpretation of the Raman modes was straightforward in terms of the crystal symmetry.

Before the detailed discussion of the results, we describe the selection rule of Raman scattering of $\alpha$-TeO$_2$. Raman tensors of the point group of $D_4$ (422) in the coordinate, in which $x$-, $y$-, and $z$-axes are parallel to [100], [010], and [001] directions of the crystal, respectively, are given by
\begin{widetext}
\begin{eqnarray}
A_1=\left(
\begin{array}{ccc}
a&0&0\\
0&a&0\\
0&0&b\\
\end{array} 
\right),\; 
B_1=\left(
\begin{array}{ccc}
c&0&0\\
0&-c&0\\
0&0&0\\
\end{array} 
\right),\; 
B_2=\left(
\begin{array}{ccc}
0&d&0\\
d&0&0\\
0&0&0\\
\end{array} 
\right),\; 
E_x=\left(
\begin{array}{ccc}
0&0&0\\
0&0&e\\
0&e&0\\
\end{array} 
\right),\; 
E_y=\left(
\begin{array}{ccc}
0&0&-e\\
0&0&0\\
-e&0&0\\
\end{array} 
\right),
\label{eq7}
\end{eqnarray}
\end{widetext}
where $a$, $b$, $c$, $d$, and $e$ are the constant \cite{ASPine}. Our Raman scattering experiments were conducted on (110)- and (101)-oriented single crystals, so that we defined new coordinates as $x^\prime$-, $y^\prime$-, and $z^\prime$-axes for experiments using a (110)-oriented crystal and as $x^{\prime\prime}$-, $y^{\prime\prime}$-, and $z^{\prime\prime}$-axes for experiments using a (101)-oriented crystal. In the new coordinate, in which $x^\prime$-, $y^\prime$-, and $z^\prime$-axes are parallel to [110], [1$\bar{1}$0], and [001] directions of the crystal, respectively, Raman tensors in Eq. (4) are converted to 
\begin{widetext}
\begin{eqnarray}
A_1^\prime=\left(
\begin{array}{ccc}
a&0&0\\
0&a&0\\
0&0&b\\
\end{array} 
\right),\; 
B_1^\prime=\left(
\begin{array}{ccc}
0&-c&0\\
-c&0&0\\
0&0&0\\
\end{array} 
\right),\; 
B_2^\prime=\left(
\begin{array}{ccc}
d&0&0\\
0&-d&0\\
0&0&0\\
\end{array} 
\right),\; 
E_{x}^{\prime}=\frac{1}{\sqrt{2}}\left(
\begin{array}{ccc}
0&0&e\\
0&0&e\\
e&e&0\\
\end{array} 
\right),\; 
E_{y}^{\prime}=\frac{1}{\sqrt{2}}\left(
\begin{array}{ccc}
0&0&-e\\
0&0&e\\
-e&e&0\\
\end{array} 
\right).
\label{eq7-1}
\end{eqnarray}
\end{widetext}
In another new coordinate, in which $x^{\prime\prime}$-, $y^{\prime\prime}$-, and $z^{\prime\prime}$-axes are parallel to [101], [010], and [10$\bar{1}$] directions of the crystal, respectively, Raman tensors in Eq. (4) are converted to 
\begin{widetext}
\begin{eqnarray}
A_1^{\prime\prime}=\frac{1}{2}\left(
\begin{array}{ccc}
a+b&0&-a+b\\
0&2a&0\\
-a+b&0&a+b\\
\end{array} 
\right),\; 
B_1^{\prime\prime}=\frac{1}{2}\left(
\begin{array}{ccc}
c&0&-c\\
0&-2c&0\\
-c&0&c\\
\end{array} 
\right),\; 
B_2^{\prime\prime}=\frac{1}{\sqrt{2}}\left(
\begin{array}{ccc}
0&d&0\\
d&0&-d\\
0&-d&0\\
\end{array} 
\right),\;\nonumber\\ 
E_{x}^{\prime\prime}=\frac{1}{\sqrt{2}}\left(
\begin{array}{ccc}
0&e&0\\
e&0&e\\
0&e&0\\
\end{array} 
\right),\; 
E_{y}^{\prime\prime}=\frac{1}{\sqrt{2}}\left(
\begin{array}{ccc}
-e&0&0\\
0&0&0\\
0&0&e\\
\end{array} 
\right).
\label{eq7-2}
\end{eqnarray}
\end{widetext}
We listed in Table II the selection rule of the Raman tensors given by Eqs. (5) and (6).

Figures 3(a) to (f) show the polarized Raman spectra in various configurations in the frequency region of 1.5--7 THz. The frequency resolution was $\sim$ 2 cm$^{-1}$, as indicated by a pair of vertical lines. In Raman spectrum in $x^\prime(z^\prime z^\prime)\overline{x^\prime}$ configuration [Fig. 3(a)], we observe a peak structure at $\sim$ 4.46 THz. According to the selection rule of Raman modes listed in Table II, only $A_1$ mode is active in this configuration. Thus, the observed mode can be assigned to the $A_1$ mode. The tiny peak structure at $\sim$ 3.71 THz, indicated by an asterisk, is due to the leakage of the Raman mode discerned in $x^\prime(y^\prime z^\prime)\overline{x^\prime}$ configuration [Fig. 3(b)]. In $x^\prime(y^\prime z^\prime)\overline{x^\prime}$ configuration, only the longitudinal optical E ($E_{\rm LO}$) mode is active (see Table II) and thus the peak structures at $\sim$ 3.71 THz and $\sim$ 5.92 THz are assigned to the $E_{\rm LO}$ modes. The tiny peak structure at $\sim$ 4.46 THz (indicated by an asterisk) arises from the leakage of $A_1$ mode observed in $x^\prime(z^\prime z^\prime)\overline{x^\prime}$ configuration [Fig. 3(a)]. In $x^\prime(y^\prime+z^\prime y^\prime+z^\prime)\overline{x^\prime}$ configuration [Fig. 3(c)], two peak structures are observed at $\sim$ 3.71 THz and $\sim$ 5.92 THz, which can be assigned to $E_{\rm LO}$ modes by comparing them with the Raman spectrum in $x^\prime(y^\prime z^\prime)\overline{x^\prime}$ configuration [Fig. 3(b)]. In addition to two $E_{\rm LO}$ modes, the peak structures at $\sim$ 4.49 THz and $\sim$ 4.65 THz are discerned. Such peak structures are more pronounced in $x^\prime(y^\prime y^\prime)\overline{x^\prime}$ configuration [Fig. 3(d)]. In this configuration, $A_1$ and $B_2$ modes are active (see Table II). Within the frequency resolution, the center frequency of the observed mode ($\sim$ 4.49 THz) in $x^\prime(y^\prime y^\prime)\overline{x^\prime}$ configuration corresponds to that of $A_1$ mode ($\sim$ 4.46 THz) in $x^\prime(z^\prime z^\prime)\overline{x^\prime}$ configuration [Fig. 3(a)]. Thus, the peak structures at $\sim$ 4.49 THz and $\sim$ 4.65 THz in $x^\prime(y^\prime y^\prime)\overline{x^\prime}$ configuration would be assigned to $A_1$ and $B_2$ modes, respectively.

To confirm the mode assignment described above, we measured Raman spectra rotating the (110)-oriented single crystal by the angle $\varphi$, as illustrated in the inset of Fig. 4(a). We defined the laboratory coordinate as $X$ (horizontal)-axis and $Y$ (vertical)-axis. $\varphi$ was defined as the angle of the [1$\overline{1}$0] direction of the crystal relative to the $X$-axis. Figures 4(a) and (b) show the $\varphi$ dependence of the integrated Raman scattering intensity (circles) in the frequency region of 1.5--7 THz in $E_{\rm i} \parallel E_{\rm s}$ and $E_{\rm i} \perp E_{\rm s}$ configurations, respectively. $E_{\rm i}$ and $E_{\rm s}$ represent the electric field of the incident and scattering light, respectively. $E_{\rm s}$ was set parallel to the $X$-axis, while $E_{\rm i}$ was set parallel to the $X$- or $Y$-axis. The integrated Raman intensity depends on $\varphi$. According to Raman tensors given in Eq. (5), the integrated Raman intensity $I$ in $E_{\rm i} \parallel E_{\rm s}$ configuration can be expressed as
\begin{equation}
I_\parallel=(a\cos^2\varphi+b\sin^2\varphi)^2+d^2\cos^4\varphi+4e^2\cos^2\varphi\sin^2\varphi.
\end{equation}
$\varphi=0^\circ$, $\varphi=45^\circ$, and $\varphi=90^\circ$ correspond to $x^\prime(y^\prime y^\prime)\overline{x^\prime}$, $x^\prime(y^\prime+z^\prime y^\prime+z^\prime)\overline{x^\prime}$, and $x^\prime(z^\prime z^\prime)\overline{x^\prime}$ configurations, respectively. On the other hand, $I$ in $E_{\rm i} \perp E_{\rm s}$ configuration can be expressed as 
\begin{eqnarray}
I_\perp &=&(-a+b)^2\cos^2\varphi\sin^2\varphi\nonumber\\
&&+d^2\cos^2\varphi\sin^2\varphi+e^2(\cos\varphi^2-\sin\varphi^2)^2.
\end{eqnarray}
$\varphi=0^\circ$ corresponds to $x^\prime(y^\prime z^\prime)\overline{x^\prime}$ configuration. The fitting results using Eqs. (7) and (8) are shown by orange lines in Figs. 4(a) and (b), respectively. The observed $\varphi$ dependences can be well reproduced by adjusting the Raman tensors of $a$, $b$, $d$, and $e$. Blue, red, and green lines indicate the estimated $\varphi$ dependence of $A_1$, $B_2$, and $E_{\rm LO}$ modes, respectively. Thus, we can conclude that two peak structures at $\sim$ 4.49 THz and $\sim$ 4.65 THz in $x^\prime(y^\prime y^\prime)\overline{x^\prime}$ configuration [Fig. 3(d)] are due to $A_1$ and $B_2$ modes, respectively. 

In $x^{\prime\prime}(z^{\prime\prime} z^{\prime\prime})\overline{x^{\prime\prime}}$ configuration [Fig. 3(e)], four peak structures emerge at $\sim$ 1.88 THz, $\sim$ 3.67 THz, $\sim$ 4.48 THz, and $\sim$ 5.26 THz, which are assigned to $B_1$, transverse optical $E$ ($E_{\rm TO}$), $A_1$, and $E_{\rm TO}$ modes, respectively (see Table II). In the $x^{\prime\prime}(y^{\prime\prime} y^{\prime\prime})\overline{x^{\prime\prime}}$ configuration [Fig. 3(f)], $B_1$ modes are more pronounced, while $E_{\rm TO}$ mode disappears. Our mode assignments based on the polarized Raman scattering experiments using (110)- and (101)-oriented single crystals are consistent with the mode assignments reported in Refs. \onlinecite{ASPine} and \onlinecite{VRodriguez}. It should be noted that the frequencies of the Raman active $E_{\rm TO}$ modes nearly coincide with the frequencies of the peak structures (3.66 THz and 5.37 THz) in the $\alpha_{\rm o}$ spectrum shown in Fig. 3(g). A detailed discussion about this is given in Sec. V. D.

\subsection{Terahertz radiation}

Figure 5(a) illustrates the experimental setup for the terahertz radiation. In this experiment, the [010] direction of a (101)-oriented crystal was set parallel to the $X$-axis. The electric field of a femtosecond laser pulse was set parallel to the $Y$-axis. By using the wire-grid polarizer in front of the LT-GaAs detector, we detected the terahertz electric field $E_{\rm THz}$ polarized along the $X$-axis, i.e., $E_{\rm THz}\parallel X$. The penetration depth at 800 nm (the wavelength of a femtosecond laser pulse used here) was estimated to be $\sim$ 6 mm [see the $\alpha$ spectrum in Fig. 2(h)], which exceeds the thickness of the sample (0.5 mm). Thus, $\alpha$-TeO$_2$ is a transparent with respect to the incident femtosecond laser pulse. By the irradiation of a femtosecond laser pulse, we found terahertz radiation from a (101)-oriented $\alpha$-TeO$_2$. The measured waveform is shown in Fig. 5(b); a single-cycle pulse with a pulse width of $\sim$ 0.3 ps is observed at around 0 ps. After the single-cycle pulse, the temporal oscillation component with a period of $\sim$ 0.27 ps is discerned up to $\sim$ 3 ps. At around 3.5 ps, a second pulse shows up and the temporal oscillation component is also observed up to $\sim$ 8 ps. To see the spectrum of the terahertz radiation, we performed the Fourier transformation of the measured waveform. The frequency resolution was estimated to be 0.01 THz, which was determined by the measured time window $\sim$ 100 ps. The obtained power spectrum of the terahertz radiation is shown in Fig. 5(c). Below 3 THz, a fringe pattern is discerned with a period of $\sim$ 0.25 THz. Furthermore, a sharp peak appears at $\sim$ 3.71 THz with a full-width at the half maximum of $\sim$ 0.1 THz. It should be noted that the sensitivity of our detection system has the maximum at $\sim$ 1.6 THz and decreases from $\sim$ 1 at 1.6 THz to $\sim$ 0.07 at 3.71 THz. This indicates that the actual generation efficiency of the terahertz radiation at 3.71 THz is larger than that below 3 THz by a factor of $\sim$ 10 as discussed later in Sec. V. B. The 3.71 THz peak corresponds to the oscillation with a period of $\sim$ 0.27 ps in Fig. 5(b). We discuss the origins of the single-cycle terahertz pulse at 0 ps in terms of the optical rectification in Sec. V. A, the intensity fringe pattern with a period of $\sim$ 0.25 THz in Sec. V. B, the second pulse at around 3.5 ps in Sec. V. C, and the temporal oscillation component with a frequency of 3.71 THz in Sec. V. D.

To estimate the efficiency of the terahertz radiation from $\alpha$-TeO$_2$, we selected, as a reference sample, ZnTe, which is known as a typical terahertz emitter. In the same experimental setup, we measured terahertz radiation from a 0.5-mm-thick (110)-oriented ZnTe crystal. The amplitude of the terahertz electric field of $\alpha$-TeO$_2$ was evaluated to be $\sim$ 1/300 of that of ZnTe. The second-order nonlinear optical coefficient of $\alpha$-TeO$_2$ is 0.59 pm/V \cite{DSChemla} or 0.69 $\pm$ 0.08 pm/V at 1.064 $\mu$m \cite{SSingh}, while that of ZnTe is $\sim$ 50--120 pm/V \cite{RKChang,HPWagner}. Thus, the observed difference of the terahertz electric fields cannot be fully explained by the difference of these values alone. In $\alpha$-TeO$_2$, there is a difference ($\sim$ 2.2) between $n$ in the terahertz frequency region [Fig. 2(c)] and $n_{\rm g}$ at 800 nm  [Fig. 2(g)]. Thus, the observed difference of the terahertz electric fields can be related to the difference of the coherence length or the effective generation length for the terahertz radiation. The detailed discussion about them is given in Sec. V. B.

\subsection{Azimuth angle and laser power dependence of terahertz radiation}

Here, we show the results of the azimuth-angle and laser-power dependences of the terahertz radiation. We measured the waveform from -0.8 ps to 2.7 ps rotating the sample by an angle $\theta$ around the propagation vector of the femtosecond laser pulse, as illustrated in Fig. 5(a). $\theta$ was defined as the angle of the [010] direction of the crystal relative to the $X$-axis. The electric field $E^\omega$ of the femtosecond laser pulse was set parallel to the $X$- or $Y$-axis. We detected the $X$-axis component of the terahertz waves ($E_{\rm THz}\parallel X$). Figures 6(a) and (b) show the $\theta$ dependence of the terahertz electric fields $E^\omega\parallel Y$ ($E_{\rm THz}^{XY}$ ) and $E^\omega\parallel X$ ($E_{\rm THz}^{XX}$), respectively. Both quantities were obtained by integrating the amplitude spectrum in the frequency range of 0--3.2 THz, which is characterized by the component of the single-cycle pulse at around 0 ps, as guided by the horizontal arrow in the upper side of Fig. 5(c). They exhibit a notable $\theta$ dependence. The absolute value of $E_{\rm THz}^{XY}$ reaches the maxima at $\sim$ 10$^\circ$ and $\sim$ 190$^\circ$, while the sign of $E_{\rm THz}^{XY}$ is reversed by rotating the sample by 180$^\circ$. On the other hand, the absolute value of $E_{\rm THz}^{XX}$ reaches the maxima at $\sim -50^\circ$ and $\sim$ 130$^\circ$ and shows the second maxima at $\sim$ 60$^\circ$ and $\sim$ 240$^\circ$. We show in Fig. 6(c) the laser power dependence of $E_{\rm THz}^{XY}$ (circles) with $\theta$ = 0$^\circ$; $E_{\rm THz}^{XY}$ linearly increases with the laser power. We discuss the observed $\theta$ and laser power dependences of terahertz radiation in Sec. V. A in terms of the optical rectification.

Figures 6(d) and (e) show the $\theta$ dependence of $E_{\rm THz}^{XY}$ and $E_{\rm THz}^{XX}$ of the oscillation component, respectively. Both quantities were obtained by integrating the amplitude spectrum in the frequency range of 3.2--5.0 THz, as indicated by the horizontal arrow in the upper part of Fig. 5(c). Figure 6(f) shows the laser power dependence of $E_{\rm THz}^{XY}$ of the oscillation component with $\theta$ = 0$^\circ$. Observed $\theta$ and the laser power dependences are almost the same as those of $E_{\rm THz}^{XX}$ and $E_{\rm THz}^{XY}$ of the single-cycle terahertz pulse shown in Figs. 6 (a)--(c). We will discuss these results in Sec. V. D.

\section{Discussion}
\subsection{Origin of single-cycle terahertz radiation at 0 ps}
Here, we explain the observed $\theta$ and laser power dependences of the single-cycle terahertz pulse at 0 ps in terms of the optical rectification described in Sec. II. $\alpha$-TeO$_2$ is known to show the birefringence in the transparent frequency region. Indeed, we observed the birefringence ($\sim$ 0.16) at 800 nm, which is characterized by the difference between $n_{\rm o}$ and $n_{\rm e}$ [see Fig. 2(g)]. Thus, we consider the effect of the birefringence on the nonlinear polarization given by Eq. (2). $P_{010}$ in Eq. (2) is not influenced by the birefringence because it is generated by the $[10\overline{1}]$ component of $E^\omega$ alone. On the other hand, $P_{10\overline{1}}$ is generated by mixing of the [010] and $[10\overline{1}]$ components of $E^\omega$, which propagate with different phase velocities because of the birefringence. As a result, $P_{10\overline{1}}$ changes its phase rapidly with the propagation of a femtosecond laser pulse. This causes the destructive interference of the terahertz waves generated within the crystal. Therefore, $P_{10\overline{1}}$ in Eq. (2) can be negligible, and the final form of $P_{\rm NL}^X$ in the $E^\omega\parallel X$ configuration becomes
\begin{equation}
P_{\rm NL}^X = P_{010} = \epsilon _{0} d_{14} E^{2}_{0}\sin^2\theta.
\label{eq3}
\end{equation}
In our experimental geometry [Fig. 5(a)], we only detected the terahertz electric field $E_{\rm THz}$ along the $X$-axis ($E_{\rm THz}^{XX}$). Therefore, $E_{\rm THz}^{XX}$ can be expressed as
\begin{equation}
E^{XX}_{\rm THz} \propto   \epsilon _{0} d_{14} E^{2}_{0}\sin^2\theta\cos\theta.
\label{eq4}
\end{equation}
In the $E^\omega\parallel Y$-axis configuration, $P_{NL}^Y$ in Eq. (3) is given by
\begin{equation}
P_{\rm NL}^Y = P_{010} = \epsilon _{0} d_{14} E^{2}_{0}\cos^2\theta.
\label{eq5}
\end{equation}
Accordingly, $E_{\rm THz}$ along the $Y$-axis ($E_{\rm THz}^{XY}$) follows
\begin{equation}
E^{XY}_{\rm THz} \propto   \epsilon _{0} d_{14} E^{2}_{0}\cos^3\theta.
\label{eq6}
\end{equation}
Using Eq. (12), we calculated $E_{\rm THz}^{XY}$ as a function of $\theta$, which is indicated by the red line in Fig. 6(a). It shows the maxima at $\theta$ = 0$^\circ$ and its sign is reversed by rotating the sample by 180$^\circ$. Although the overall trend of the measured $E_{\rm THz}^{XY}$ (indicated by open circles) is consistent with the calculated $E_{\rm THz}^{XY}$, the $\theta$ dependence of the measured $E_{\rm THz}^{XY}$ is totally shifted by $\sim$ 10$^\circ$. We also calculated $E_{\rm THz}^{XX}$ values using Eq. (10) and plotted them in Fig. 6(b). The measured $E_{\rm THz}^{XX}$ values [open circles in Fig. 6(b)] also shift by $\sim$ 10$^\circ$ as compared to the calculated values.	

Next, we discuss this discrepancy by taking into account the optical activity of $\alpha$-TeO$_2$ at 800 nm. $\alpha$-TeO$_2$ is known to show the optical activity for light propagating along the optic axis [19], in which the polarization of light rotates during the propagation ($\sim$ 48.5$^\circ$/mm at 800 nm). In our experimental geometry [Fig. 5(a)], the femtosecond laser pulse propagates along the [101] direction, which is inclined by 45$^\circ$ from the optic axis. The sample thickness used here was 0.5 mm. Thus, the rotation of the polarization of a femtosecond laser pulse was estimated to be $\sim$ 10$^\circ$ [$\sim$ 48.5$^\circ$/mm $\times$ 0.5mm $\times$ (cos45$^\circ$)$^2$]. Black line in Fig. 6(a) indicates the result of the least-square fitting of $E_{\rm THz}^{XY}$ by taking into account the effect of the rotation of the polarization ($\sim$ 10$^\circ$). The fitting result well reproduces the $\theta$ dependence of the measured $E_{\rm THz}^{XY}$. The $\theta$ dependence of $E_{\rm THz}^{XX}$ is also reproduced as indicated by the black line, as shown in Fig. 6(b). From Eq. (12), $E_{\rm THz}^{XY}$ is expected to increase linearly with the laser power ($E_0^2$). This relation was confirmed for $E_{\rm THz}^{XY}$ (circles) with $\theta=0^\circ$ in Fig. 6(c). From these results, we can conclude that the single-cycle terahertz pulse at 0 ps comes from the optical rectification described by Eqs. (10) and (12).

\subsection{Intensity fringe pattern in power spectrum}

First, we consider the frequency dependence of the detection sensitivity of terahertz electric fields in our system. We calculated the instrumental function $H_{\rm inst}$ by considering the response function of the LT-GaAs detector when an ideal terahertz waveform was introduced. The calculation procedure of $H_{\rm inst}$ is detailed in Appendix C. As shown in Fig. 7(a), $H_{\rm inst}$ has a peak structure at $\sim$ 1.6 THz and extends to $\sim$ 4.0 THz.

To reveal the origin of the intensity fringe pattern with a period of $\sim$ 0.25 THz in the power spectrum of terahertz radiation [Fig. 5(c)], we calculated the coherence length $l_{\rm c}$ for the terahertz radiation. $l_{\rm c}$ is defined as 
\begin{equation}
l_{\rm c}=\frac{\lambda_{\rm THz}}{2|n_{\rm g}-n_{\rm THz}|},
\end{equation}
where $\lambda_{\rm THz}$ is the terahertz wavelength and $n_{\rm THz}$ is the refractive index in the terahertz frequency region \cite{ANahata}. In the calculation of $l_{\rm c}$, we used the value of $n_{\rm o}$ in the terahertz frequency region, since the electric field of the detected terahertz wave is parallel to the ordinary ray direction, as discussed in Sec. V. A [see Eqs. (9) and (11)]. $n_{\rm g}$ at 800 nm was estimated to be $\sim$ 2.52 [Fig. 2(g)], which is indicated by the horizontal broken line in Fig. 2(c). The calculated $l_{\rm c}$ as a function of frequency is shown in Fig. 7(b). The $l_{\rm c}$ spectrum shows the small peak structure at around 3.70 THz, yielding $l_{\rm c}$ $\sim$ 28 $\mu$m. With lowering frequency, $l_{\rm c}$ gradually increases and reaches $\sim$140 $\mu$m at 0.5 THz. Such frequency dependence is due to the decrease of $n_{\rm o}$ with lowering frequency and the resultant decrease of the difference between $n_{\rm o}$ in the terahertz frequency region and $n_{\rm g}$ at 800 nm, as seen in Fig. 2(c).

At 3.66 THz, a strong absorption assigned to the $E_{\rm TO}$ mode exists [Fig. 2(d)]. To take into account the effect of this absorption, we calculated the effective generation length for the terahertz radiation $L_{\rm gen}$ (Ref. \onlinecite{ASchneider}), which represents the length of the region effectively contributing to the observed terahertz waves. Using the absorption coefficient $\alpha_{\rm o}$ shown in Fig. 2(d), $L_{\rm gen}$ is expressed as
\begin{widetext}
\begin{equation}
L_{\rm gen}=\left(\frac{1+\exp(-\alpha_{\rm o} d)-2\exp\left(-\frac{\alpha_{\rm o}}{2} d \right)\cos\left(\frac{\omega}{c}\mid n_{\rm o}-n_{\rm g}\mid d\right)}{\left(\frac{\alpha_{\rm o}}{2}\right)^2+\left(\frac{\omega}{c}\right)^2(n_{\rm o}-n_{\rm g})^2}\right)^{1/2}.
\end{equation}
\end{widetext}
We show in Fig. 7(c) the calculated $L_{\rm gen}$ as a function of frequency. Because of the presence of the $E_{\rm TO}$ mode, $L_{\rm gen}$ is reduced in the wide frequency range down to 0.2 THz; $L_{\rm gen}$ at around 3.7 THz becomes $\sim$ 8 $\mu$m. Furthermore, $L_{\rm gen}$ oscillates in the frequency region of 0.5--3 THz. The period of this oscillation was estimated to be $\sim$ 0.25 THz, which is determined by the cosine term of Eq. (14) and is in good agreement with that of the intensity oscillation pattern observed in the power spectrum of the terahertz radiation shown in Fig. 7(e). Thus, we can conclude that the observed intensity fringe pattern originates from the poor phase-matching condition with $|n_{\rm o}-n_{\rm g}|$ $\sim$ 2.2 below 3 THz. 

Finally, we obtained the frequency dependence of the terahertz radiation efficiency in our system, which was derived from $H_{\rm inst}\times L_{\rm gen}^2$, as shown in Fig. 7(d). The estimation procedure of the instrumental function $H_{\rm inst}$ is detailed in Appendix C. The calculated spectrum of the terahertz radiation efficiency can reproduce well the intensity fringe pattern in the measured power spectrum of the terahertz radiation shown in Fig. 7(e). Although the peak structure at 3.71 THz is observed in Fig. 7(d), its magnitude is much smaller than that of the intensity fringe pattern below 3 THz. This fact indicates that another mechanism is necessary to explain the peak structure at 3.71 THz. We will discuss its origin in Sec. V. D.

\subsection{Second pulse at around 3.5 ps}

Here, we discuss the reason why the second pulse appears at around 3.5 ps in the waveform [Fig. 5(b)]. A possible reason for the generation of a second pulse is a multiple reflection of the incident femtosecond laser pulse or the radiated terahertz pulse between the sample-air interfaces. In these cases, we can expect the interval ($\Delta t$) between the single-cycle pulse at 0 ps and the second pulse using the relation $\Delta t= 2n_{\rm g}d/c$ or $\Delta t=2n_{\rm o}d/c$ and the value of $n_{\rm g}$ ($\sim$ 2.52) at 800 nm [Fig. 2(g)] or $n_{\rm o}$ ($\sim$ 4.6) below 2 THz [Fig. 2(c)]. Here, $c$ and $d$ are the speed of light and the thickness of the sample (0.5 mm), respectively. The values thus expected are $\Delta t$ $\sim$ 7.0 ps and $\Delta t$ $\sim$ 15.3 ps for the multiple reflection of the femtosecond laser pulse and of the radiated terahertz pulse, respectively. The observed interval ($\sim$ 3.5 ps) does not accord with the calculated $\Delta t$ values. In our experimental condition, the magnitude of $l_{\rm c}$ or $L_{\rm gen}$ is the order of 10--100 $\mu$m [Figs. 7(c) and (b)], which is several times shorter than the thickness of the sample (0.5 mm) used here. This poor phase-matching condition for the terahertz radiation results in the emergence of two terahertz pulses; we reproduced the measured terahertz waveform by frequency- and time-domain calculations, as detailed in Appendix D. In this case, the interval of two terahertz pulses is expressed as $\Delta t=|n_{\rm o}-n_{\rm g}|d/c$ and equal to $\sim$ 3.5 ps, which agrees well with the observed $\Delta t$. Thus, the single-cycle terahertz pulse at around 0 ps and the second pulse at around 3.5 ps can be assigned to terahertz waves generated in the rear and front regions of the crystal, respectively, as illustrated in Fig. 5(a).

\subsection{Peak structure at 3.71 THz in power spectrum}
In this section, we discuss the origin of the sharp peak structure at 3.71 THz in the power spectrum of the terahertz radiation. A narrowband terahertz radiation can be generated by the increase of $l_{\rm c}$ or $L_{\rm gen}$ in a narrow frequency region, where the phase-matching condition ($n_{\rm g}\sim n_{\rm THz}$) in Eq. (13) is satisfied. Indeed, the temporal oscillation component in the terahertz wave has been observed near the phonon-polariton resonance in ZnTe \cite{CMTu} and the $F_2$ phonon mode in Bi$_4$Ge$_3$O$_{12}$ \cite{RTakeda}. In $\alpha$-TeO$_2$, the calculated generation efficiency of the terahertz radiation shows the peak structure at 3.71 THz [Fig. 7(d)]. However, its intensity is more than three orders of magnitude smaller than that below 3 THz. Thus, the increase of $l_{\rm c}$ or $L_{\rm gen}$ cannot explain the observed power spectrum of terahertz radiation shown in Fig. 7(e).

A possible origin of the peak structure at 3.71 THz is the enhancement of the $\chi^{(2)}$ near the phonon mode. Considering the symmetry of the $\chi^{(2)}$ tensor given by Eq. (1), the induced $P_{\rm NL}$ is always parallel to the ordinary ray direction. Thus, $P_{\rm NL}$ induced by the optical rectification is coupled only with infrared active phonon modes for ordinary ray, i.e., the $E_{\rm TO}$ mode \cite{DMKorn}. Figure 8(a) shows the magnified view of $\alpha$ spectrum (the green line) [see Fig. 2(d)] and the loss-function --Im [1/$\epsilon$] spectrum (the red line), which were calculated by using fitting parameters listed in Table III (see Appendix B). We also show in Fig. 8(b) the polarized Raman spectra in $x^{\prime\prime}(z^{\prime\prime} z^{\prime\prime})\overline{x^{\prime\prime}}$ and $x^\prime(y^\prime z^\prime)\overline{x^\prime}$ configurations, which represent $E_{\rm TO}$ and $E_{\rm LO}$ modes, respectively. Within the frequency resolution ($\sim$ 2 cm$^{-1}$) indicated by a pair of vertical lines, the peak positions of $E_{\rm TO}$ and $E_{\rm LO}$ modes in the $\alpha$ spectrum are identical to those in Raman spectrum, respectively. Apparently, the peak position of $E_{\rm TO}$ mode in the $\alpha$ and Raman spectra is different from the peak position of the power spectrum of the terahertz radiation shown in Fig. 8(c). This clearly excludes the possibility of the enhancement of the $\chi^{(2)}$ as an origin of the peak structure at 3.71 THz in the power spectrum of the terahertz radiation.

Another possible origin of the 3.71 THz radiation is the coherent phonon generation of the $E$ mode by ISRS \cite{YXYan,TPDougherty}. To discuss this possibility, we first examine the conservation law of the wavevector dominating the ISRS mechanism. By the irradiation of a femtosecond laser pulse to the sample, Raman active modes are coherently excited. By taking into account the conservation law of the wavevector, the coherent phonon should be generated with the wavevector $q = 0$, i.e., LO mode. As can be seen in Fig. 8(c), the peak position of the power spectrum of terahertz radiation just corresponds to the peak position of $E_{\rm LO}$ mode in the $\alpha$ spectrum [the red line in Fig. 8(a)] and Raman spectrum in $x^\prime(y^\prime z^\prime)\overline{x^\prime}$ configuration [the red circles in Fig. 8(b)]. This clearly indicates that terahertz radiation by coherent phonons occurs by ISRS mechanism.

In ISRS mechanism of terahertz radiation, the related phonon modes are both Raman and infrared active. In the measured window of the terahertz radiation, $A_1$, $A_2$, $B_1$, $B_2$, and $E$ modes exist, as seen in optical and Raman spectra shown in Figs. 3(a)--(f) and (g), respectively. $A_1$, $B_1$, and $B_2$ modes show no effect on the generation process of the terahertz waves, because infrared-inactive modes cannot be coupled with $P_{\rm NL}$ \cite{YRShen}. $A_2$ mode has also no contribution to the terahertz radiation, since it is Raman inactive. Thus, only the $E$ mode contributes to the terahertz radiation. Here, we discuss the role of Raman process in terms of the symmetry analysis and explain the observed $\theta$ dependence of $E_{\rm THz}^{XY}$ and $E_{\rm THz}^{XX}$ of the oscillation component, as shown in Figs. 6(d) and (e), respectively. We simply consider the three level systems; the ground, middle, and excited levels are denoted as $\ket{1}$, $\ket{2}$, and $\ket{3}$, respectively, as illustrated in the inset of Fig. 8(c). The matrix element in the process from $\ket{1}$ to $\ket{2}$ through $\ket{3}$ is given by $M_{jk}=\braket{2|\mu_j|3}\braket{3|\mu_k|1}$, where $M_{jk}$ represents the Raman activity of the mode. On the other hand, the matrix element in the process from $\ket{2}$ to $\ket{1}$, is represented by $A_i=\braket{1|\mu_i|2}$. In the case of the $E_x$ mode, $M_{jk}^x$ is given by $E_x$ in Eq. (4). The dipole-moment is polarized along the $x$-axis, thus $A_i^x=(\beta, 0, 0)$. In this case, the tensor of $A_i^x M_{jk}^x$ using the contradicted notation becomes
\begin{eqnarray}
\begin{array}{ccc}
A_{i}^xM_{jk}^x & = & 
\left[
\begin{array}{cccccc}
0 & 0 & 0 & \beta e & 0 &0 \\
0 & 0 & 0 & 0 & 0 &0 \\
0 & 0 & 0 & 0 & 0 &0 \\
\end{array}
\right].
\end{array}
\end{eqnarray}
In the case of the $E_y$ mode, $M_{jk}^y$ is given by $E_y$ in Eq. (4). The dipole-moment is polarized along the $y$-axis, thus $A_i^y=(0, \beta, 0)$. In this case, the tensor of $A_i^y M_{jk}^y$ using the contradicted notation becomes
\begin{eqnarray}
\begin{array}{ccc}
A_i^yM_{jk}^y & = & 
\left[
\begin{array}{cccccc}
0 & 0 & 0 & 0 & 0 &0 \\
0 & 0 & 0 & 0 & -\beta e &0 \\
0 & 0 & 0 & 0 & 0 &0 \\
\end{array}
\right].
\end{array}
\end{eqnarray}
Thus, the tensor of $A_i^xM_{jk}^x+A_i^yM_{jk}^y$ is given by
\begin{eqnarray}
\begin{array}{ccc}
A_i^xM_{jk}^x+A_i^yM_{jk}^y & = & 
\left[
\begin{array}{cccccc}
0 & 0 & 0 & \beta e & 0 &0 \\
0 & 0 & 0 & 0 & -\beta e &0 \\
0 & 0 & 0 & 0 & 0 &0 \\
\end{array}
\right].
\end{array}
\end{eqnarray}
Non-zero tensor components of $A_i^xM_{jk}^x+A_i^yM_{jk}^y$ are identical to that of $\chi^{(2)}$ in Eq. (1). Therefore, the $\theta$ dependence of the terahertz electric field induced by ISRS is the same as that induced by optical rectification, as we observed here [Figs. 6(d) and (e)]. This selection rule is different from the well-known selection rule determined by the conventional reflection- or transmission-type pump-and-probe experiments. These experiments measure the change of $n$ via the modulation of $R$ and/or $T$. In this case, the modulation signal should obey the selection rule determined by the Raman tensor. This is because that the induced $n$ is proportional to $R_{jkl} E_k^\omega E_l^\omega$, where $R_{jkl}$ is the Raman tensor \cite{TDekorsyREV}. Indeed, it was confirmed that the angle dependence of the modulation signal was the same as that of the Raman tensor in an optically transparent insulator LaAlO$_3$ \cite{YLiu} and even in an optically opaque GaAs/AlGaAs multiple quantum wells \cite{KJYee}. However, in the terahertz radiation experiments, the modulation of $n$ alone cannot radiate the terahertz wave into free space. Therefore, the $\theta$ dependence of the terahertz electric field is different from that of the Raman tensor, being contrast to the case of the pump-and-probe experiments \cite{YLiu,KJYee}.

\section{Summary}

In summary, we observed terahertz radiation at room temperature in a non-centrosymmetric and insulating oxide, paratellurite $\alpha$-TeO$_2$, by the irradiation of a femtosecond laser pulse. The intensity fringe pattern with a period of $\sim$ 0.25 THz appeared in the power spectrum of the radiated terahertz wave. This intensity fringe pattern was reproduced by taking into account the frequency dependence of the effective generation length for the terahertz radiation with a poor phase-matching condition. Furthermore, we observed a temporal oscillation component in the radiated terahertz wave with a frequency of $\sim$ 3.71 THz, which just corresponds to the center frequency of the Raman active longitudinal optical $E$ mode. On the basis of the comprehensive polarized optical and Raman spectroscopic studies, we concluded that the impulsive stimulated Raman scattering (ISRS) generates the coherent phonon of the longitudinal optical $E$ mode, resulting in the emission of the terahertz waves at 3.71 THz.

\begin{acknowledgments}
We thank A. Doi, J. Fujioka, and Y. Tokura for their support in far-infrared reflectance measurements. This work was partly supported by Murata Foundation and by a Grant-in-Aid by MEXT (No. 25247049, 25247058, and No. 25-3372). M. S. was supported by Japan Society for the Promotion of Science (JSPS) through Program for Leading Graduate Schools (MERIT) and JSPS Research Fellowships for Young Scientists.
\end{acknowledgments}

\appendix
\section{Estimation of complex optical constants in terahertz frequency region by terahertz time-domain spectroscopy}

Terahertz time-domain spectroscopy obtains the amplitude and phase spectra by the fast Fourier transformation (FFT) of the measured terahertz waveforms with and without the sample. The black line in Fig. 9 shows the terahertz waveform generated from a (110)-oriented ZnTe crystal, which was used as a reference. The transmitted terahertz waveforms passing through the sample in $E^\omega\parallel [001]$ and $E^\omega\parallel [1\overline{1}0]$ configurations are indicated by blue and red lines, respectively. The amplitude of the transmitted terahertz waveforms decreases due to the absorption of the sample. Their phases are also delayed due to the refractive index of the sample. The power transmission $T$ and phase $\phi$ spectra can be obtained by the following relationship; 
\begin{equation}
\sqrt{T}\exp({-i\phi})\equiv\frac{E_{\rm sample}}{E_{\rm reference}},
\end{equation}
where $E_{\rm sample}$ and $E_{\rm reference}$ are FFT spectra with and without the sample, respectively. The obtained $T$ spectra in $E^\omega\parallel [001]$ and $E^\omega\parallel [1\overline{1}0]$ configurations are shown by circles and squares in Fig. 2(b), respectively.

Without taking into account the multiple reflections inside the sample, $E_{\rm sample} / E_{\rm reference}$ in Eq. (A1) is expressed as
\begin{equation}
\frac{E_{\rm sample}}{E_{\rm reference}}=\frac{2}{\tilde{n}+1}\frac{2\tilde{n}}{\tilde{n}+1}\exp\left(-i\frac{\omega}{c}d(\tilde{n}-1)\right),
\end{equation}
where $\omega$ is the frequency, $c$ is the velocity of light, $d$ is the thickness of the sample, and $\tilde{n} (= n + i\kappa)$ is the complex refractive index of the sample. We avoided the effect of the multiple reflections by restricting the time range of the FFT. We numerically derived the complex optical spectra from Eq. (A2) using experimentally determined $T$ and $\phi$.

\section{Estimation of complex optical constants in terahertz frequency region by Fourier-transformed infrared spectroscopy}

Figure 10(a) shows the polarized reflectance $R$ spectra in the energy range of 7--60 meV, measured by Fourier-transformed infrared (FT-IR) spectroscopy. $R$ spectra in $E^\omega\parallel [001]$ and $E^\omega\parallel [1\overline{1}0]$ configurations are shown by circles and squares, respectively. In order to extract complex optical constants in the terahertz frequency region, we performed the dispersion analysis of the measured $R$ spectra. Real $\epsilon_1(\omega)$ and imaginary $\epsilon_2(\omega)$ parts of the dielectric constant, composed of $N$ Lorentz oscillators are, respectively, given by
\begin{equation}
\epsilon _{1}(\omega) = \epsilon_{\infty} +\sum_{i=1}^N \left[\frac{f_i\omega_i^2(\omega_{i}^2-\omega^2)} {(\omega_{i}^2-\omega^2)^2+\gamma_i^2\omega^2} \right],
\end{equation}
\begin{equation}
\epsilon _{2}(\omega) = \sum_{i=1}^N \left[ \frac{f_i\omega_i^2\gamma_i\omega} {(\omega_{i}^2-\omega^2)^2+\gamma_i^2\omega^2} \right],
\end{equation}
where $\epsilon_\infty$, $\gamma_i$, $\omega_i$, and $f_i$ represent the high frequency dielectric constant, the damping coefficient, the central frequency, and the oscillator strength, respectively. The measured $R$ spectra in the energy range of 7--100 meV were reproduced by the fitting parameters listed in Table III, as indicated by the solid lines in Fig. 10(a). We assumed four and eight Lorentz oscillators in $E^\omega\parallel [001]$ and $E^\omega\parallel [1\overline{1}0]$ configurations, respectively. With the use of the estimated parameters, we calculated the refractive index $n$ and absorption coefficient $\alpha$ spectra for ordinary and extraordinary rays, which are shown in Figs. 10(b) and (c), respectively. The peak energies of the observed modes nearly correspond to the peak energies of the modes reported previously (Ref. \onlinecite{DMKorn}). According to the mode symmetry, $A_2$ modes are infrared active for extraordinary ray, i.e., $E^\omega\parallel [001]$, while $E$ modes are infrared active for ordinary ray, i.e., $E^\omega\parallel [1\overline{1}0]$. Thus, the observed modes for ordinary and extraordinary rays can be assigned to $E$ and $A_2$ modes, respectively. As seen in the magnified $n$ and $\alpha$ spectra shown in Figs. 2(c) and (d), $n$ and $\alpha$ spectra obtained by terahertz time-domain spectroscopy in the transmission geometry are nearly the same as the $n$ and $\alpha$ spectra calculated by the procedure described above. This ensures the validity of the present dispersion analysis.

\section{Estimation of instrumental function}

We calculated the instrumental function in our experimental setup by considering the response function of the detector when an ideal terahertz wave of optical rectification process was introduced. We used a low-temperature-grown GaAs (LT-GaAs) detector coupled with a dipole-antenna. The detector characteristics may be governed by the resonance properties of the antenna structure and by the carrier lifetime \cite{SGPark}. However, it was reported that the response function of the LT-GaAs detector is independent to the resonance character of the antenna \cite{SKono} and the gap size of the antenna \cite{SMatsuura}. Thus, we excluded the effect of the antenna structure and simply assumed the response function of the LT-GaAs by taking into account the pulse width of the femtosecond laser pulse and the carrier lifetime of the LT-GaAs. The time profile of the femtosecond laser pulse was ideally given by the following Gaussian intensity envelope; 
\begin{equation}
I(t)=I_0\exp(-\frac{2t^2}{\tau^2}),
\end{equation}
where $\tau$ is the pulse width of a femtosecond laser pulse used here ($\tau$ = 100 fs). The time profile of the conductivity of the LT-GaAs $\sigma(t)$ induced by delta-function excitation is simply expressed by
\begin{eqnarray}
\sigma(t)&=&\sigma_0\exp(-\frac{t}{\tau_{\rm c}})\hspace*{1cm} (t>0),\nonumber\\
&=&0\hspace*{2.7cm} (t\leq 0),
\end{eqnarray}
where $\tau_{\rm c}$ is the lifetime of the photogenerated carriers. For calculation, we used $\tau_{\rm c}$ = 0.3 ps, which was estimated by the pump-and-probe reflection measurements reported in Ref. \onlinecite{MTaniLT}. The calculated $\sigma(t)$ is shown by the dashed line in Fig. 11(a). The response function of the detector in time domain $H(t)$ can be obtained by convolution of $\sigma(t)$ and $I(t)$: 
\begin{equation}
H(t)=\int^\infty_{-\infty}dt^\prime I(t^\prime)\sigma(t-t^\prime),
\end{equation}
which is shown by the red line in Fig. 11(a). The amplitude of the response function $H(\omega)$ is given by
\begin{equation}
H(\omega)=\int^\infty_{-\infty}dt\exp(-i\omega t)\int^\infty_{-\infty}dt^\prime I(t^\prime)\sigma(t-t^\prime).
\end{equation}
We also included the term of $\omega$ as the terahertz waves spread by diffraction at the detector position. Since we can focus the terahertz waves in the diffraction limited size, the detected amplitude is inversely proportional to the wavelength. Thus, we used $\omega H(\omega)$ as the detection response function, which is shown by the red line in Fig. 11(b). The phase of $\omega H(\omega)$ is uniquely determined by the condition, i.e., $\displaystyle \lim_{\omega \rightarrow 0} \arg [H(\omega)]=0$ and is plotted by the blue line in Fig. 11(b). Thus, the intensity response function of the LT-GaAs ($I_{\rm res}$) is proportional to $\omega^2|H(\omega)|^2$, which is shown by the blue line in Fig. 11(c). $I_{\rm res}$ reaches the maxima at $\sim$ 0.6 THz and extends to $\sim$ 4 THz.

We further consider the spectrum of the ideal terahertz wave, which is introduced into the LT-GaAs detector. The ideal terahertz wave is proportional to the second time derivative of the nonlinear polarization $P_{\rm NL}$. The induced $P_{\rm NL}$ would be proportional to the shape of a femtosecond laser pulse. Thus, we simply assumed that the ideal terahertz wave can be represented by the second time derivative of the Gaussian intensity envelope given by Eq. (1). The calculated spectrum of the ideal terahertz wave $I_{\rm THz}(\omega)$ is shown by the black line in Fig 11(c). It shows the peak structure at $\sim$ 2.3 THz. Then, we obtained the instrumental function $I_{\rm inst}$ in our experimental setup, which was derived from $I_{\rm res}\times I_{\rm THz}$. The red line in Fig. 11(d) shows the calculated $I_{\rm inst}$. 

To confirm the validity of the calculated $I_{\rm inst}$, we measured the waveform of the terahertz radiation emitted from 0.5-mm-thick (110)-oriented ZnTe and 0.2-mm-thick (110)-oriented GaP crystals in the same experimental setup. $l_{\rm c}$ of ZnTe and GaP are the order of $\sim$ mm \cite{JFNye}, exceeding the sample thickness used here. Indeed, the second-order nonlinear susceptibility $\chi^{(2)}$ spectrum in ZnTe is nearly constant below 3 THz \cite{SMatsuura}. Thus, the spectral shapes of terahertz radiation of ZnTe and GaP are mainly determined by $I_{\rm inst}$. We show in Fig. 11(d) the Fourier-transformed spectra of the measured waveforms of ZnTe and GaP, which are indicated by the green and purple lines, respectively. These spectra are quite resemble to the calculated $I_{\rm inst}$ indicated by the red line, except for the reduction of the power spectrum of ZnTe above 3 THz. This reduction is due to the presence of the longitudinal acoustic phonon mode at $\sim$ 3.7 THz with $\alpha$ $\sim$ 90 cm$^{-1}$ \cite{GGallot}.

\section{Waveform simulations}

Here, we discuss the origin of the second pulse at around 3.5 ps in the observed terahertz waveform with $E^\omega\parallel Y$ and $\theta=0^\circ$, as shown in Fig. 12(a). In order to isolate the temporal oscillation component with a frequency of 3.71 THz, we performed the inverse Fourier transformation of the power spectrum [Fig. 5(c)] in the frequency range of 3.2--5 THz and obtained the waveform of the oscillation component, which is shown in Fig. 12(b). The amplitude oscillation before -0.5 ps is an artifact due to the abrupt change of the amplitude at 3.2 THz. We also extracted the waveform of the single component by the inverse Fourier transformation of the power spectrum [Fig. 5(c)] in the frequency range of 0--3.2 THz, as shown in Fig. 12(c). We can discern two terahertz pulses; first pulse at around 0 ps and second pulse at around 3.5 ps. In order to discuss the origin of the second pulse, here we simulated the terahertz waveforms in $\alpha$-TeO$_2$ by both frequency- and time-domain calculations using optical constants in the terahertz and visible frequency regions.

\subsection{Waveform simulation by frequency-domain calculation}

In the frequency-domain simulation, we solved a wave equation of the terahertz electric field. Optical rectification process of the femtosecond laser pulse produces nonlinear polarization $P_{\rm NL}$ \cite{ASchneider}:
\begin{equation}
P_{\rm NL}(\omega)=\frac{\chi^{\rm OR}(\omega, \omega_{\rm opt})}{n(\omega_{\rm opt})c}I_0(\omega),
\end{equation}
where, $\omega_{\rm opt}$ is the optical carrier frequency, $\chi^{\rm OR}(\omega, \omega_{\rm opt})=\chi^{\rm (2)}(\omega; \omega_{\rm opt}, \omega-\omega_{\rm opt})$ is the second-order nonlinear optical susceptibility, and $I_0(\omega)$ is the Fourier transformed intensity spectrum of the femtosecond laser pulse. Laser-induced $P_{\rm NL}(\omega)$ becomes the wave source for the terahertz radiation. Thus, the wave equation of the terahertz electric field in the plane-wave approximation is expressed by \cite{ASchneider}
\begin{widetext}
\begin{equation}
\frac{\partial^2 E_{\rm THz}(\omega)}{\partial z^2}+(\frac{\omega^2 n^2(\omega)}{c^2}-i\omega\mu_0\sigma(\omega))E_{\rm THz}(\omega)=-\omega^2\mu_0\frac{\chi^{\rm OR}(\omega, \omega_{\rm opt})}{n(\omega_{\rm opt})c}
\exp(-i\frac{\omega n_{\rm g}}{c}z)\exp(-\alpha_{\rm opt}z)I_0(\omega),
\end{equation}
\end{widetext}
where $\mu_0$ is permeability in vacuum, $\sigma(\omega)=\alpha(\omega)n(\omega)/\mu_0c$ is the optical conductivity, $\alpha_{\rm opt}$ is the absorption coefficient at the frequency of the femtosecond laser pulse, $n_{\rm g}$ is the group refractive index of the femtosecond laser pulse, and $z$ is the depth from the front surface of the crystal. Equation (D2) can be solved with the following conditions; $E(\omega, z=0)=0$ and $\alpha(\omega)\ll(\omega n_{\rm g})/c$ \cite{ASchneider}. We also considered the effect of the diffraction of the terahertz waves at the illuminated area of the femtosecond laser pulse. Then, $E_{\rm THz} (\omega, z)$ in the far position outside the crystal is given by
\begin{widetext}
\begin{equation}
E_{\rm THz} (\omega, z)=\frac{\mu_0\chi^{\rm OR}(\omega, \omega_{\rm opt})\omega^2I(\omega)}{n_{\rm opt}\left(\frac{c}{\omega}(\frac{\alpha(\omega)}{2}+\alpha_{\rm opt})+i(n(\omega)+n_{\rm g})\right)}\frac{\exp(-i\frac{\omega n(\omega)}{c}z)\exp(-\frac{\alpha(\omega)}{2}z)-\exp(-i\frac{\omega n_{\rm g}}{c}z)\exp(-\alpha_{\rm opt}z)}{\frac{\alpha(\omega)}{2}-\alpha_{\rm opt}+\frac{i\omega}{c}(n(\omega)-n_{\rm g})}.
\end{equation}
\end{widetext}
The radiated terahertz waves in $\alpha$-TeO$_2$ are invariably polarized parallel to the ordinary ray direction, as detailed in Section V. A. Thus, we used the optical constants for ordinary ray, which were obtained by the reflectivity spectrum [Fig. 2(a)]. The optical constants for the $[10\overline{1}]$-polarized laser pulse at 1.55 eV are as follows; $n_{\rm g}=2.52$, $n_{\rm opt}=2.34$, and $\alpha_{\rm opt}\cong 0$ cm$^{-1}$ [Figs. 2(g) and (h)]. If the non-resonant second-order nonlinear optical process is dominant in the terahertz radiation, i.e., $\chi^{\rm OR} (\omega, \omega_0)$ is frequency-independent, we can numerically obtain the complex $E_{\rm THz} (\omega, z)$. The delay time and the sign of the waveform are arbitrary in this simulation. We further take into account the Fresnel reflection loss and the instrumental function as detailed in Appendix C. Figure 12(d) shows the simulated terahertz waveform obtained by the inverse Fourier transformation of $E_{\rm THz} (\omega, z)$. It consists of two terahertz pulses; first pulse at around 0 ps and second pulse at around 3.5 ps. The width and shape of first pulse in the simulated terahertz waveform are in good agreement with those in the measured terahertz waveform shown in Fig. 12(c). Although the shape of the second pulse can be reproduced by the simulation, the magnitude of the amplitude is slightly different. This mainly arises from the difficulty to precisely extract the absorption coefficients in the terahertz frequency region by reflectivity measurements.

\subsection{Waveform simulation by time-domain calculation}

In order to further discuss the origin of the second pulse at around 3.5 ps, we performed time-domain calculation. In this calculation, we assumed that the terahertz waves are generated at each region labeled by g$i$h along the depth position, as schematically shown in Fig. 13(a). The ideal terahertz waveform by optical rectification is expressed by
\begin{equation}
E_{\rm THz}(t)=E_0(1-\frac{4t^2}{\tau^2})\exp(-\frac{2t^2}{\tau^2}),
\end{equation}
where $E_0$ is amplitude at $t$ = 0 ps. Figure 13(b) shows the ideal terahertz waveform used in the time-domain calculation. At the rear surface ($z = d$), amplitude spectrum of the terahertz waves generated at $z$ position $E_{\rm THz} (\omega, z)$ eventually becomes 
\begin{widetext}
\begin{equation}
E_{\rm THz} (\omega, z)=E_{\rm THz}(\omega)\exp(-i\frac{\omega n_{\rm g}}{c}z)\exp(-\alpha_{\rm opt}z)\exp(-i\frac{\omega n(\omega)}{c}(d-z))\exp(-\frac{\alpha(\omega)}{2}(d-z)).
\end{equation}
\end{widetext}
Considering the diffraction of the terahertz waves, the waveform of the terahertz waves generated at the depth of $z$ in the free space is given by 
\begin{equation}
E_{\rm THz} (t, z)=\int^\infty_{-\infty}d\omega\exp(i\omega t)\omega E_{\rm THz}(\omega, z).
\end{equation}

Figure 13(c) shows the simulated terahertz waveforms at various depth positions by taking into account the Fresnel reflection at the crystal-air interface. The number of divisions $N$ was set to be 1000, thus the depth of the $i$ th division is given by $z_i =id/N$. The waveform of the terahertz wave emitted at $z = d$ ($i = N = 1000)$ is almost the same as the ideal waveform [Fig. 13(b)]. Since a femtosecond laser pulse propagates faster than the terahertz wave ($n_{\rm o}>n_{\rm g}$) [Figs. 2(c) and (g)], the phase of the terahertz wave emitted at each region is delayed; for example, the terahertz wave emitted at $z = 0$ ($i = 1)$ is delayed by $\sim$ 4 ps, compared to that emitted at $z = d$. Furthermore, the waveform is distorted due to the dispersion of $n(\omega)$ and $\alpha(\omega)$ in the terahertz frequency region. We summed the $N$ waveforms and obtained the total amplitude waveform: $E_{\rm THz}^{\rm total} (t)  =\sum_{i=1}^N E_{\rm THz}^i(t)=\sum_{i=1}^N E_{\rm THz}(t, \frac{zi}{N})$. The simulated waveform is shown in Fig. 13(d), which reproduces the characteristics of the experimentally obtained terahertz waveform in Fig. 13(e). Since the terahertz waves generated within the crystal interfere destructively except for the terahertz waves emitted in the rear and front region, the single-cycle pulse at around 0 ps and the second pulse at around 3.5 ps are mainly attributed to the terahertz waves generated at the rear and front region, respectively, which is schematically shown in Fig. 5(a).

\begin{figure}[t]
\begin{center}
\includegraphics[width=0.46\textwidth]{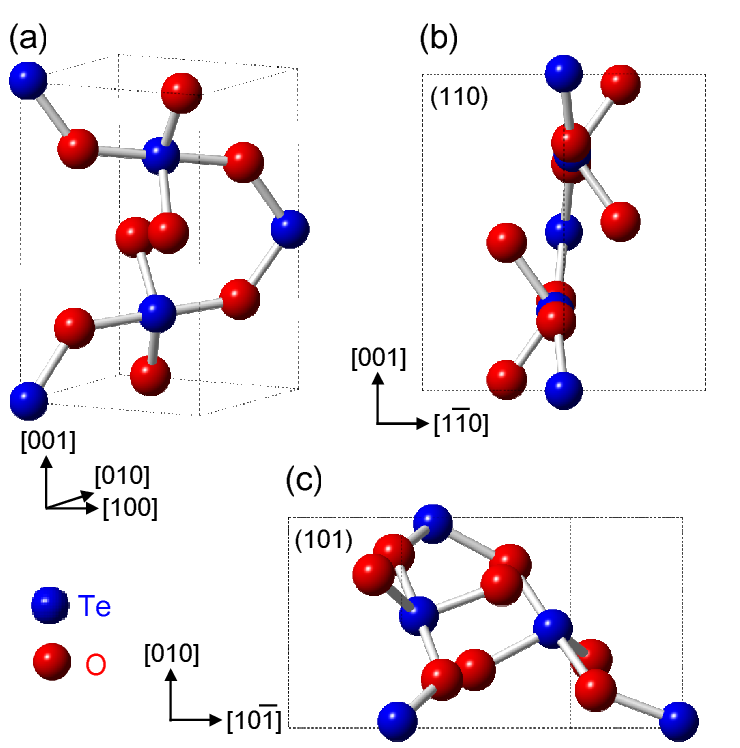}
\end{center}
\vspace*{-0.5cm} 
\caption{(Color online) Schematic illustrations of the crystal structure of $\alpha$-TeO$_2$ \cite{RWGWyckoff}. (a) Overview of the crystal structure. Crystal views of (b) (110) plane and (c) (101) plane. The dotted lines indicate the unit cell.}
\end{figure}

\begin{figure*}[t]
\begin{minipage}[t]{.62\textwidth}
\includegraphics[width=1\textwidth]{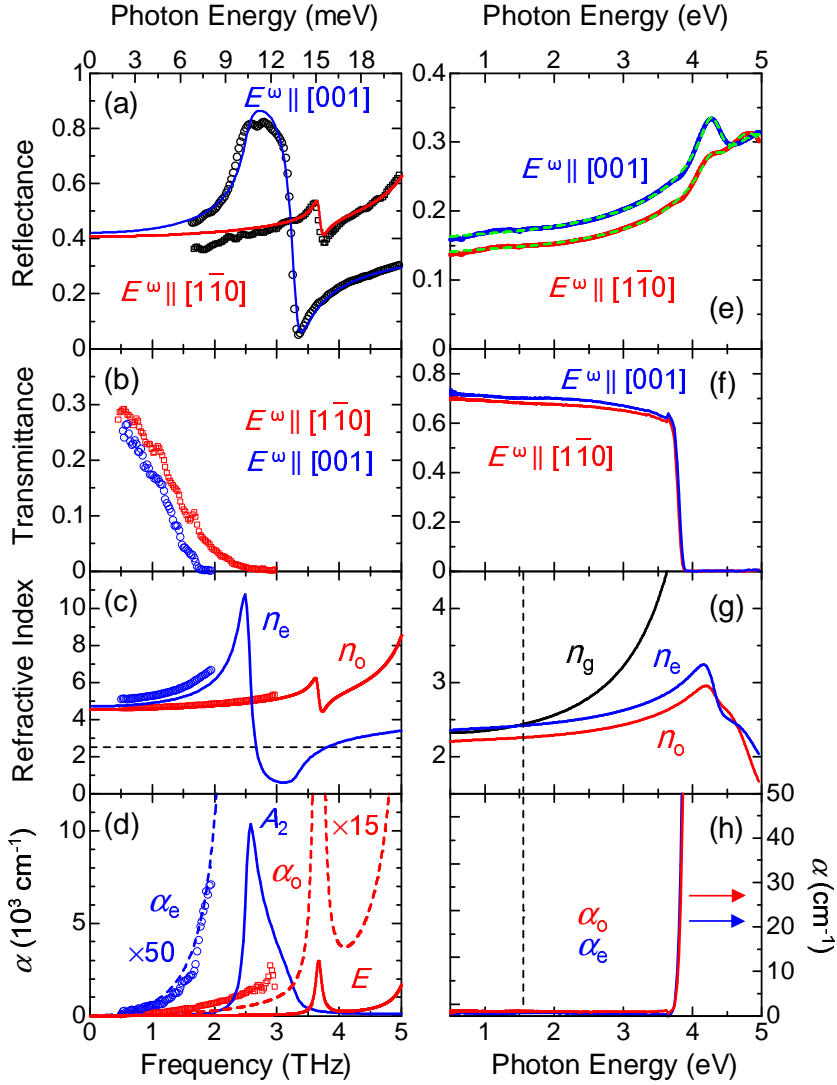}
\end{minipage}
\hfill
\begin{minipage}[b]{.33\textwidth}
\vspace*{-11cm}
\caption{(Color online) Polarized optical spectra of a (110)-oriented single crystal of $\alpha$-TeO$_2$ in terahertz, visible, and ultraviolet frequency regions. (a) Reflectance $R$ spectra in $E^\omega\parallel [001]$ (circles) and $E^\omega\parallel [1\overline{1}0]$ (squares) configurations, measured by Fourier-transformed infrared (FT-IR) spectroscopy. The solid lines are results of least-square fit to the $R$ spectra with fitting parameters listed in Table III in Appendix B. Transmittance $T$ spectra in $E^\omega\parallel [001]$ (circles) and $E^\omega\parallel [1\overline{1}0]$ (squares) configurations, measured by terahertz time-domain spectroscopy. (c) Refractive index $n$ and (d) absorption coefficient $\alpha$ spectra for ordinal ray (red) and for extraordinary ray (blue). $n$ and $\alpha$ indicated by symbols and lines were determined by terahertz time-domain spectroscopy and the dispersion analysis of the $R$ spectra, respectively. The horizontal broken line in (c) indicates the value of the group refractive index at 800 nm. (e) $R$ and (f) $T$ spectra in $E^\omega\parallel [001]$ (blue lines) and $E^\omega\parallel [1\overline{1}0]$ (red lines) configurations in 0.5--5 eV, measured by grating spectroscopy. The broken lines in (e) indicate the fitting results of the dispersion analysis with parameters listed in Table I. (g) $n$ and (h) $\alpha$ spectra for ordinal ray (red line) and for extraordinary ray (blue line). The black line in (g) represents the group refractive index $n_{\rm g}$ spectrum. The vertical broken lines in (g) and (h) indicate the photon energy of the femtosecond laser (1.55 eV) used for the terahertz radiation experiments.}
\end{minipage}
\end{figure*}

\clearpage

\begin{figure}[t]
\begin{center}
\includegraphics[width=0.44\textwidth]{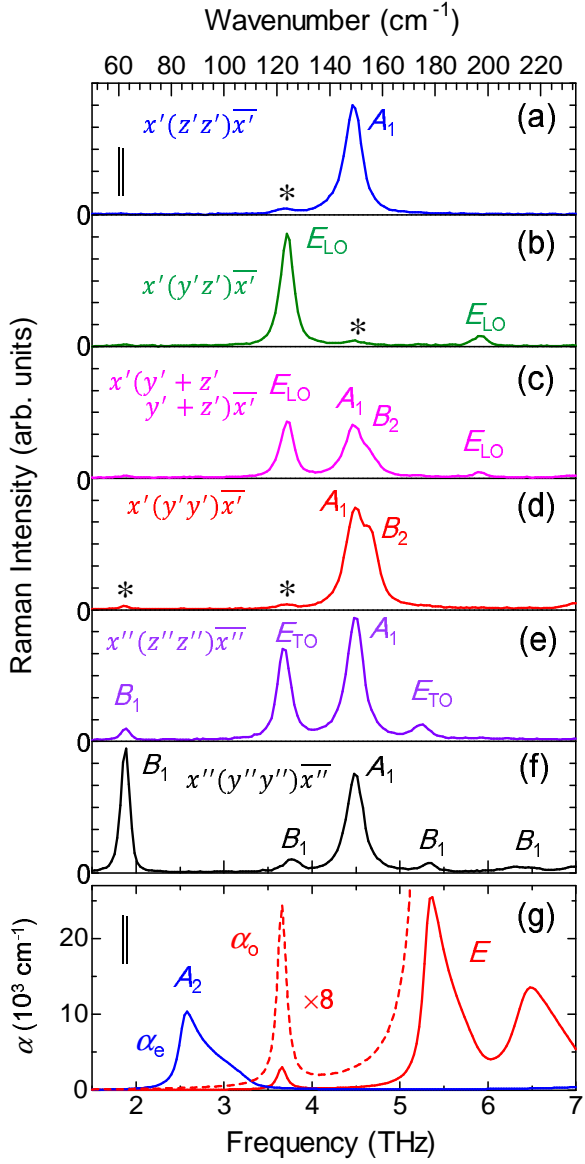}
\end{center}
\vspace*{-0.5cm} 
\caption{(Color online) Polarized Raman and optical spectra of $\alpha$-TeO$_2$ in the terahertz frequency region. (a)-(f) Raman spectra in various configurations indicated in the upper part of each spectrum. Asterisks represent the leakage of the polarization. (g) Absorption coefficient $\alpha$ spectra for ordinary and extraordinary rays. A pair of vertical lines in (a) and (g) indicate the frequency resolution.}
\end{figure}

\clearpage

\begin{figure}[t]
\begin{center}
\includegraphics[width=0.49\textwidth]{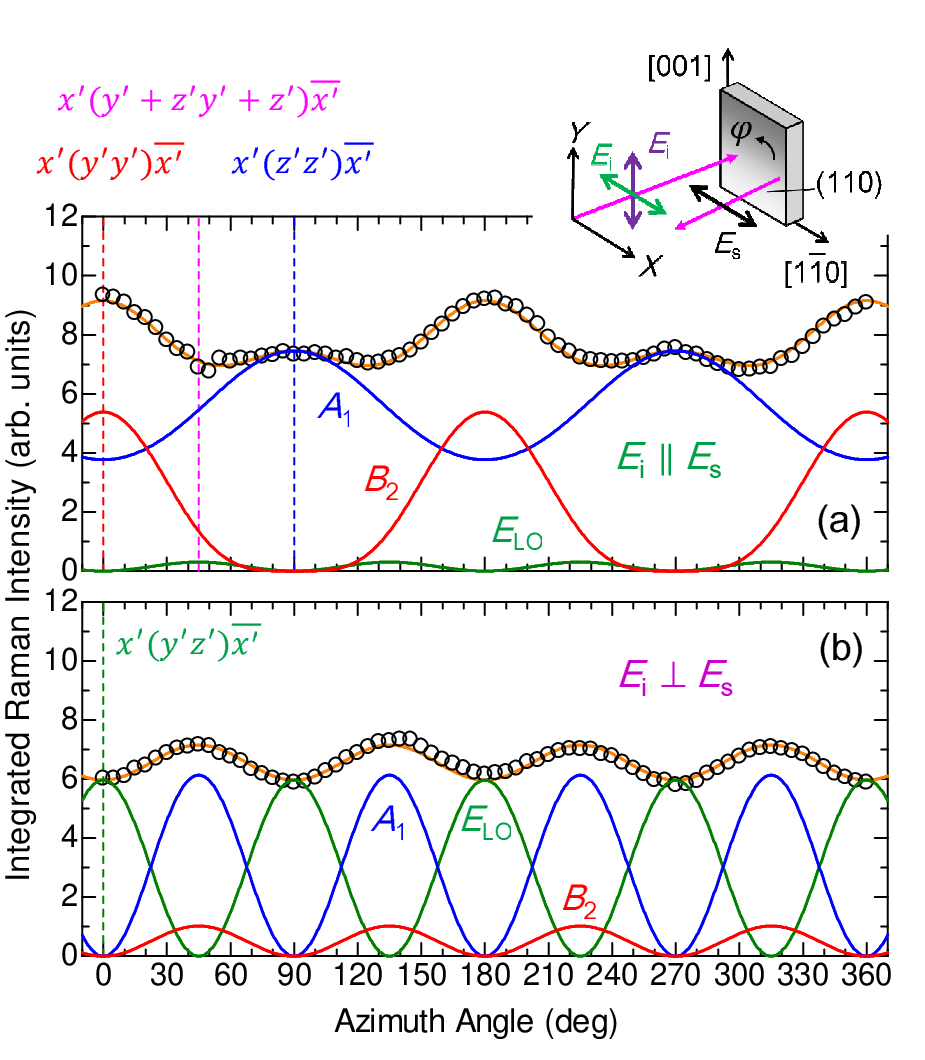}
\end{center}
\vspace*{-0.5cm} 
\caption{(Color online) Raman scattering characteristics of $\alpha$-TeO$_2$. Azimuth angle $\varphi$ dependence of the integrated Raman scattering intensity (circles) in the frequency region 1.5--7 THz in (a) $E_{\rm{i}} \parallel E_{\rm{s}}$ and (b) $E_{\rm{i}} \perp  E_{\rm{s}}$ configurations. Inset of (a) shows the schematic illustration of the experimental geometry using a (110)-oriented single crystal. $E_{\rm i}$ and $E_{\rm s}$ indicate the electric field of the incident and detected light, respectively. Orange lines are obtained from the selection rule given in Eqs. (7) and (8). $A_1$, $B_2$, and $E_{\rm LO}$ components are indicated by blue, red, and green lines, respectively.}
\end{figure}

\begin{figure*}[t]
\begin{center}
\includegraphics[width=0.95\textwidth]{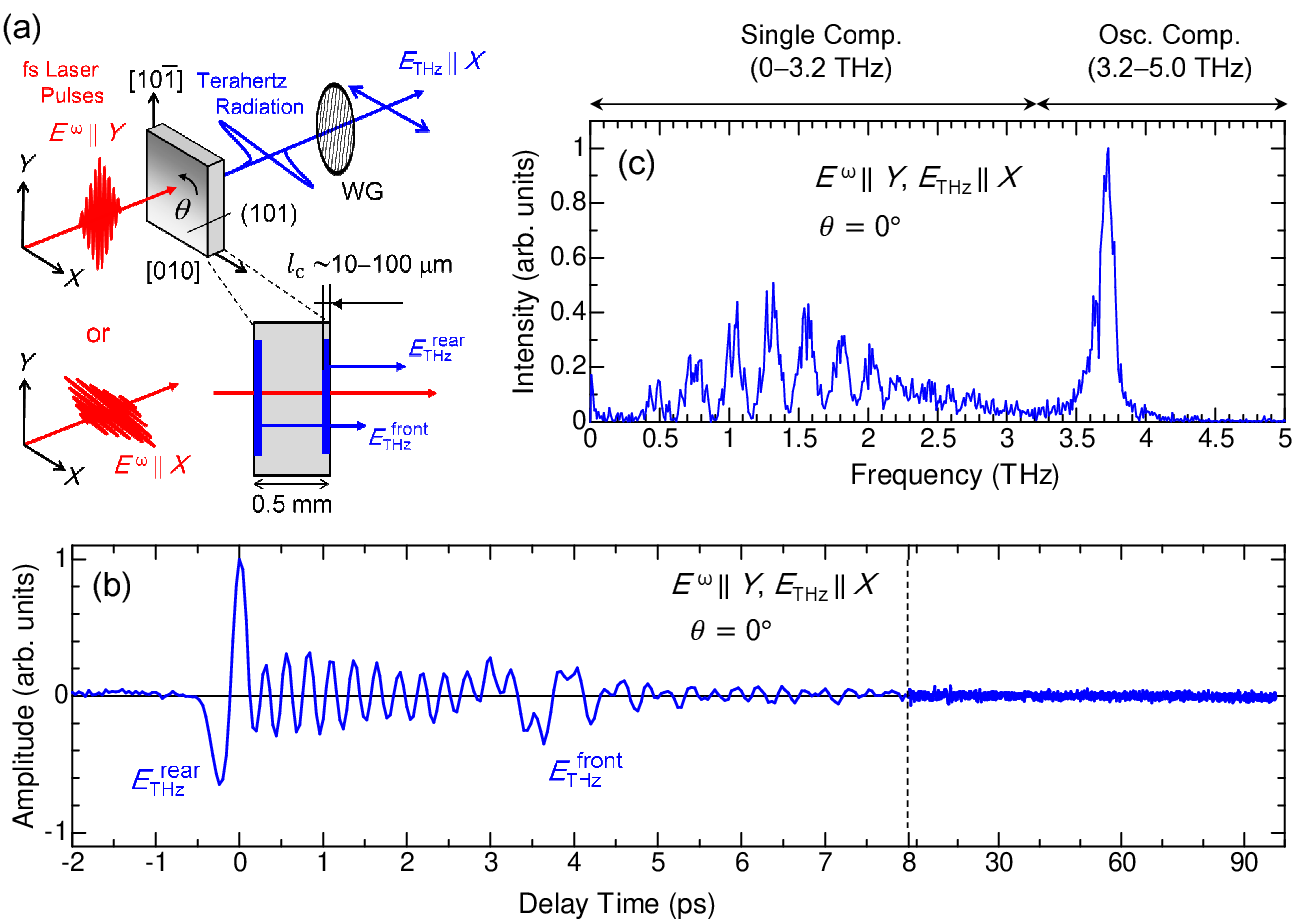}
\end{center}
\vspace*{-0.5cm} 
\caption{(Color online) Terahertz radiation from a (101)-oriented single crystal of $\alpha$-TeO$_2$ by the irradiation of a femtosecond laser pulse. (a) Experimental setup for the terahertz radiation. (b) Radiated electromagnetic wave with $E^\omega\parallel Y$ and $\theta=0^\circ$. (c) Power spectrum of the measured terahertz wave in (b).}
\end{figure*}

\begin{figure*}[t]
\begin{center}
\includegraphics[width=0.99\textwidth]{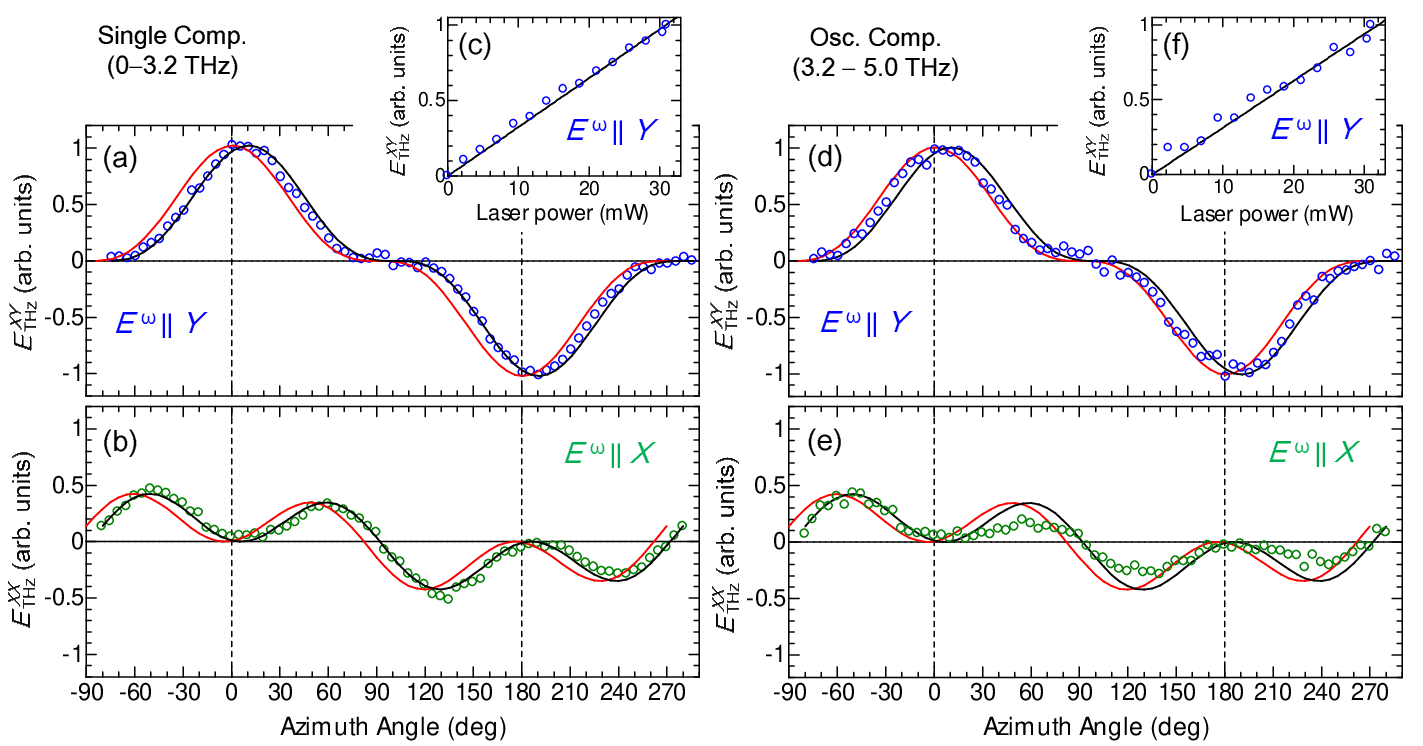}
\end{center}
\caption{(Color online) Terahertz radiation characteristics of $\alpha$-TeO$_2$. Azimuth angle dependence of the terahertz electric field $E_{\rm THz}$ (circles) of the single-cycle pulse in (a) $E^\omega\parallel Y$ ($E^{XY}_{\rm THz}$) and (b) $E^\omega\parallel X$ ($E^{XX}_{\rm THz}$). $E^{XY}_{\rm THz}$ and $E^{XX}_{\rm THz}$ were obtained by integrating the amplitude spectrum in the frequency range of 0--3.2 THz. (c) Laser power dependence of $E^{XY}_{\rm THz}$. Azimuth angle dependence of (d) $E^{XY}_{\rm THz}$ and (e) $E^{XX}_{\rm THz}$ $E_{\rm THz}$ (circles) of the temporal oscillation component, which were obtained by integrating the amplitude spectrum in the frequency range of 3.2--5.0 THz. (f) Laser power dependence of $E^{XY}_{\rm THz}$. The red lines represent the calculated curves using Eq. (12) for $E^{XY}_{\rm THz}$ and Eq. (10) for $E^{XX}_{\rm THz}$, while the black lines represent the calculated curves including the effect of the optical activity (see text).}
\end{figure*}

\clearpage

\begin{figure}[t]
\begin{center}
\includegraphics[width=0.43\textwidth]{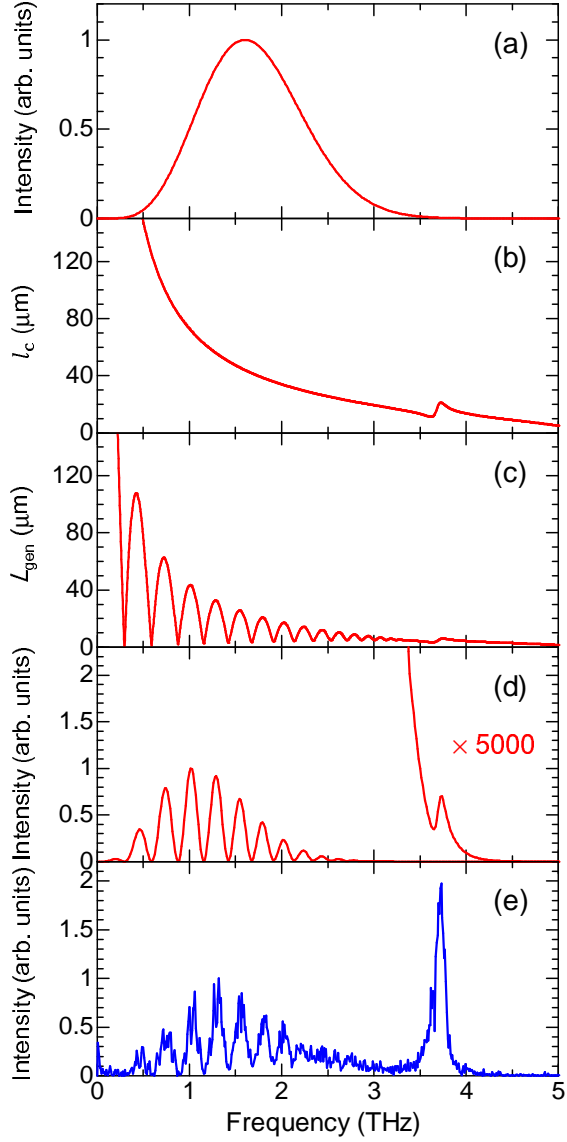}
\end{center}
\vspace*{-0.5cm} 
\caption{(Color online) Measured and calculated power spectra of terahertz radiation of $\alpha$-TeO$_2$. (a) Instrumental function of our experimental setup (see Appendix C) (b) Coherence length $l_{\rm c}$ as a function of frequency. (c) Effective generation length $L_{\rm gen}$ for the terahertz radiation as a function of frequency. (d) Calculated and (e) measured power spectra of terahertz radiation.}
\end{figure}

\clearpage

\begin{figure}[t]
\begin{center}
\includegraphics[width=0.48\textwidth]{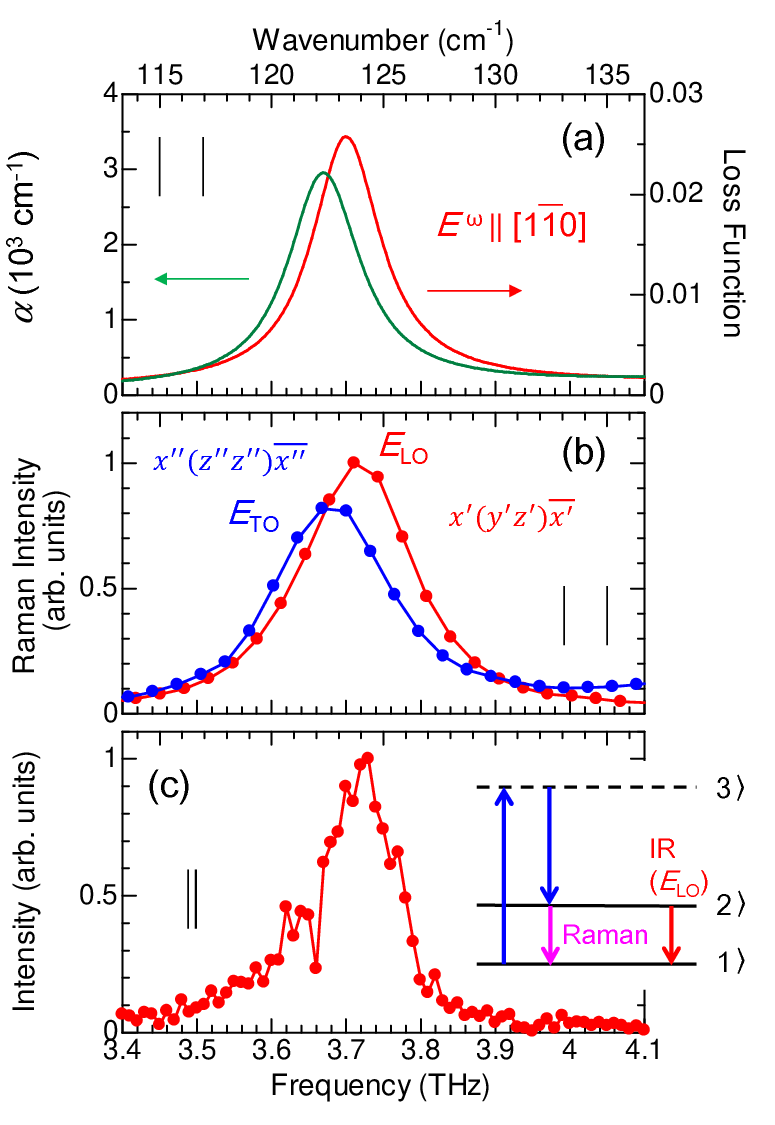}
\end{center}
\vspace*{-0.5cm} 
\caption{(Color online) Comparison of the optical spectra of $\alpha$-TeO$_2$ around 3.7 THz with the power spectrum of terahertz radiation. (a) Absorption coefficient $\alpha$ spectrum (green line) and loss function spectrum (red line). (b) Polarized Raman spectra in $x^{\prime\prime}(z^{\prime\prime} z^{\prime\prime})\overline{x^{\prime\prime}}$ and $x^\prime(y^\prime z^\prime)\overline{x^\prime}$ configurations, indicated by blue and red circles, respectively. (c) Power spectrum of terahertz radiation. A pair of vertical lines indicates the frequency resolution. Inset of (c) shows the schematic of the energy level.}
\end{figure}

\clearpage

\begin{figure}[t]
\begin{center}
\includegraphics[width=0.48\textwidth]{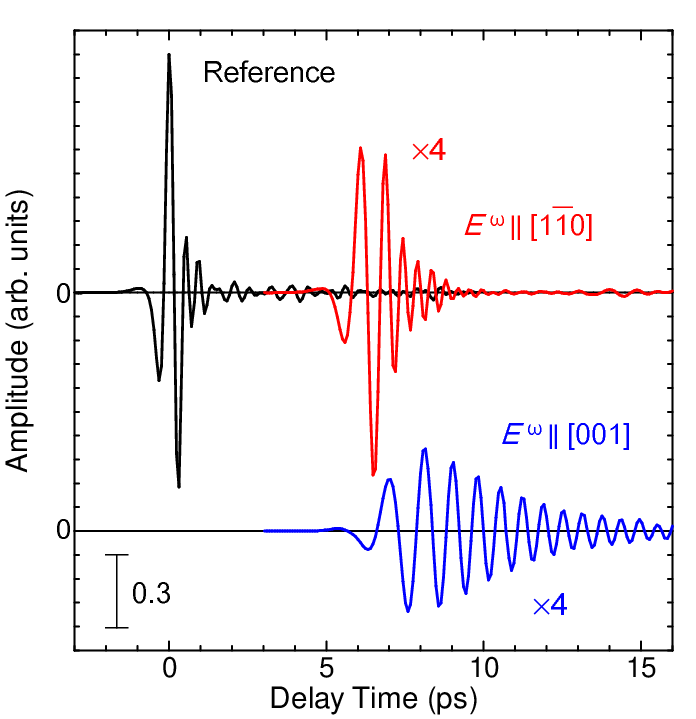}
\end{center}
\vspace*{-0.5cm} 
\caption{(Color online) Terahertz waveforms of $\alpha$-TeO$_2$ measured by terahertz time-domain spectroscopy. The reference waveform obtained from a 0.5-mm-thick (110)-oriented ZnTe crystal is shown by the black line. The transmitted terahertz waves passing through the sample in $E^\omega\parallel [001]$ and $E^\omega\parallel [1\overline{1}0]$ configurations are indicated by blue and red lines, respectively.}
\end{figure}

\clearpage

\begin{figure}[t]
\begin{center}
\includegraphics[width=0.48\textwidth]{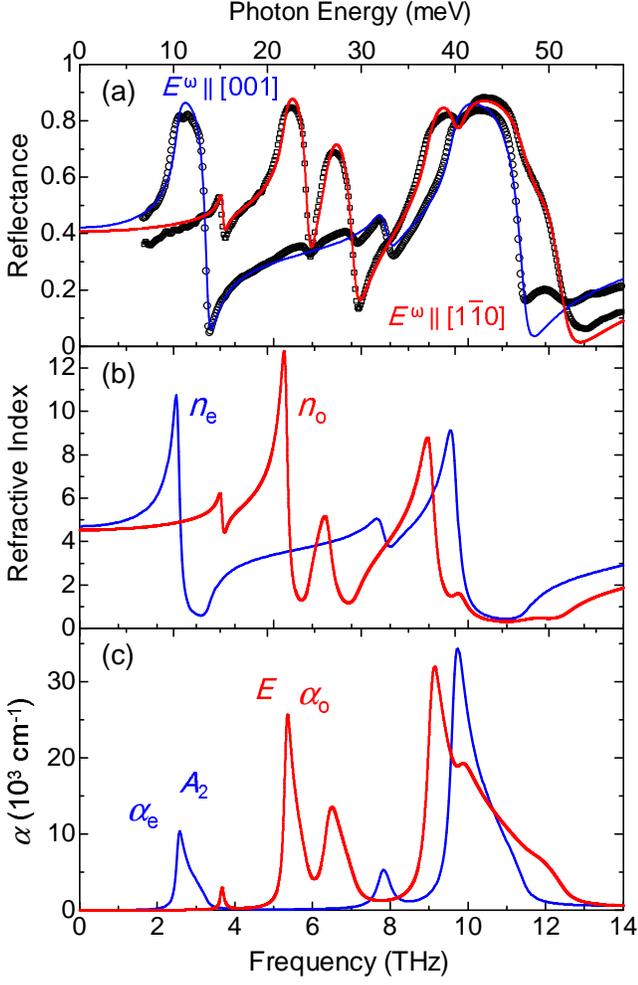}
\end{center}
\vspace*{-0.5cm} 
\caption{(Color online) Far-infrared optical spectra of $\alpha$-TeO$_2$ measured by Fourier-transformed infrared (FT-IR) spectroscopy. (a) Reflectance spectra in $E^\omega\parallel [001]$ and $E^\omega\parallel [1\overline{1}0]$ configurations are indicated by circles and squares, respectively. The solid lines are the result of least-square fit to the reflectance spectra with parameters listed in Table III. (b) Refractive index $n$ and (c) absorption coefficient $\alpha$ spectra for ordinary and extraordinary rays.}
\end{figure}

\clearpage

\begin{figure}[t]
\begin{center}
\includegraphics[width=0.49\textwidth]{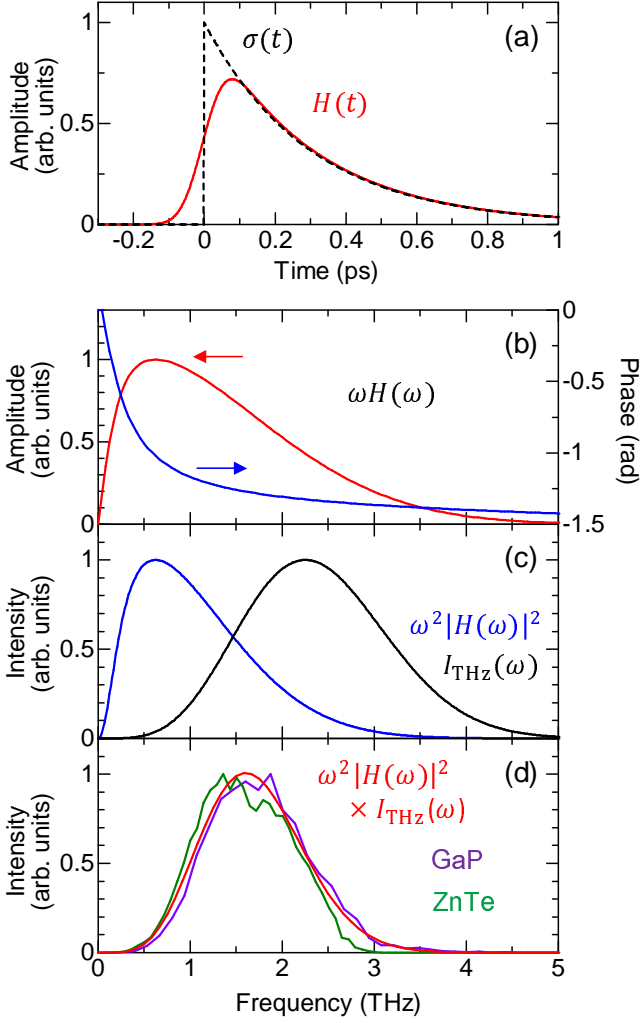}
\end{center}
\vspace*{-0.5cm} 
\caption{(Color online) Comparison of the instrumental function in our experimental setup with the power spectra of terahertz radiation in ZnTe and GaP. (a) Response function (red line) of the low-temperature-grown GaAs (LT-GaAs) detector in time domain calculated by Eq. (C3) and the photogenerated conductivity (dashed line) calculated by Eq. (C2). (b) Amplitude of the detection response function $\omega H(\omega)$ (red line) and the phase of $\omega H(\omega)$ (blue line). (c) Power spectrum of $I_{\rm THz}(\omega)$ (black line) and $I_{\rm res}(\omega) (\propto \omega^2|H(\omega)|^2)$ (blue line). (d) Red line shows the instrumental function when the ideal terahertz wave was introduced into the LT-GaAs detector. We also show the power spectra of the terahertz radiation emitted from 0.5-mm-thick (110)-oriented ZnTe and 0.2-mm-thick (110)-oriented GaP single crystals, which are shown by the green and purple lines, respectively.}
\end{figure}

\clearpage

\begin{figure}[t]
\begin{center}
\includegraphics[width=0.49\textwidth]{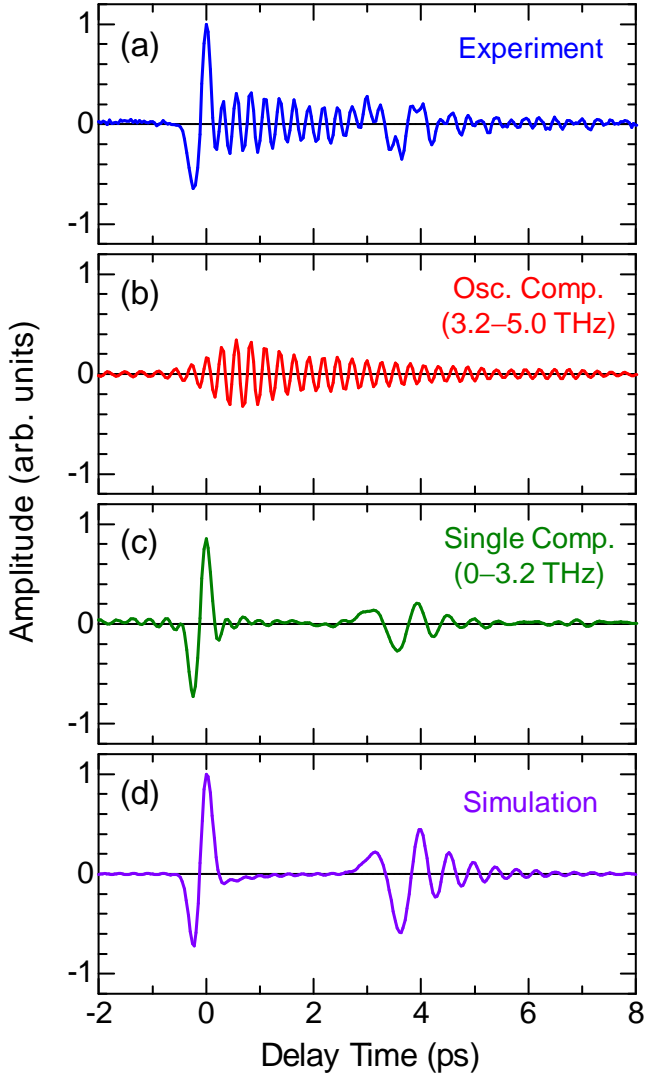}
\end{center}
\vspace*{-0.5cm} 
\caption{(Color online) Comparison of the observed terahertz waveform of $\alpha$-TeO$_2$ with $E^\omega\parallel Y$ and $\theta=0^\circ$ with the simulated terahertz waveform by frequency-domain calculation. (a) Observed terahertz waveform. (b) Temporal oscillation component with a frequency of $\sim$ 3.71 THz and (c) single-cycle component, which were derived by the inverse Fourier transformation of the power spectrum in 3.2--5.0 THz and 0--3.2 THz, respectively. (d) Terahertz waveform simulated by the frequency-domain calculation.}
\end{figure}

\clearpage

\begin{figure}[ht]
\begin{center}
\includegraphics[width=0.49\textwidth]{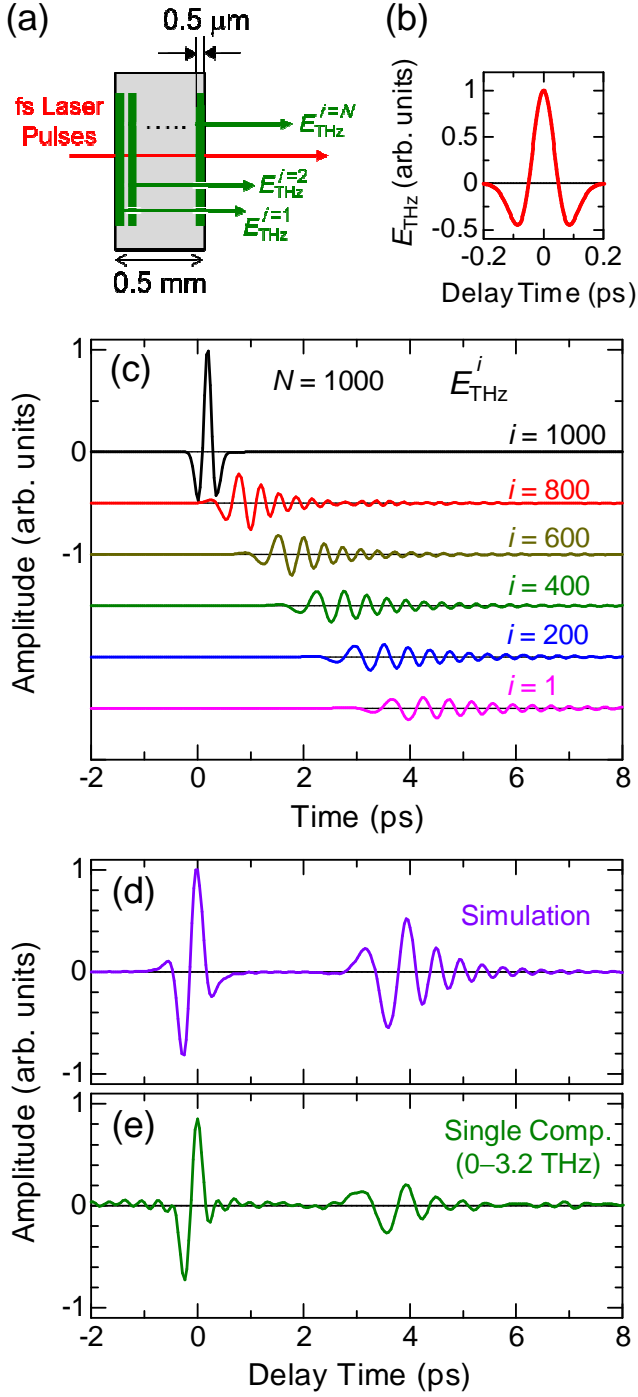}
\end{center}
\vspace*{-0.5cm} 
\caption{(Color online) Comparison of the observed terahertz waveform of $\alpha$-TeO$_2$ with $E^\omega\parallel Y$ and $\theta=0^\circ$ with the terahertz waveform simulated by time-domain calculation. (a) Schematic illustration of the waveform simulation in time domain. (b) Waveform of the ideal terahertz pulse induced by optical rectification. (c) Simulated waveforms of the terahertz waves generated at various depth positions. $i = 1$ and $i = 1000$ correspond to the front ($z = 0$) and rear ($z = 0.5$ mm) regions of the crystal, respectively. (d) Terahertz waveform simulated by the time-domain calculation. (e) Observed waveform of the single-cycle component (0--3.2 THz).}
\end{figure}

\clearpage

\begin{table}[t] 
\caption{Obtained fitting parameters of $\alpha$-TeO$_2$ in $E^\omega\parallel [001]$ configuration with the high frequency dielectric constant $\epsilon_\infty=3.47$ and $E^\omega\parallel [1\bar{1}0]$ configuration with $\epsilon_\infty=3.42$.}
    \begin{tabular}{ccccc}
      Mode $i$ & Configuration &$\omega_i$ (eV) & $\gamma_i$ (eV) & $f_i$ \\ \hline
      \hline
      1 & $E^\omega\parallel [001]$ &4.274 & 0.344 & 0.374 \\
      2 &   &4.851 & 1.308 & 1.767 \\\hline
      3 & $E^\omega\parallel [1\bar{1}0]$ &4.291 & 0.345 & 0.159 \\
      4 &   &4.701 & 0.919 & 1.336\\\hline
\end{tabular}
\end{table}

\begin{table*}[bt] 
\caption{Selection rules of infrared and Raman modes and the summary of the mode assignment of $\alpha$-TeO$_2$ in the terahertz frequency region.}
    \begin{tabular}{cccccccccccc}
     Selection Rule & Configurations & Representations & $A_1$ (THz) & $A_2$ (THz) & $B_1$ (THz) & $B_2$ (THz) & $E_{\rm TO}$ (THz) & $E_{\rm LO}$ (THz)\\ \hline
      \hline
      Raman & $x^\prime(z^\prime z^\prime)\overline{x^\prime}$ &$A_1$ & 4.46 & -- & -- & -- & -- & --\\
        & $x^\prime(y^\prime y^\prime)\overline{x^\prime}$ &$A_1+B_2$ & 4.49 & -- & -- & 4.65 & -- & --\\
        & $x^\prime(y^\prime z^\prime)\overline{x^\prime}$ &$E_{\rm LO}$ & -- & -- & -- & -- & -- & 3.71, 5.92\\
        & $x^{\prime\prime}(z^{\prime\prime} z^{\prime\prime})\overline{x^{\prime\prime}}$ &$A_1+B_1+E_{\rm TO}$ & 4.48 & -- & 1.88 & -- & 3.67, 5.26 & --\\
        & $x^{\prime\prime}(y^{\prime\prime} y^{\prime\prime})\overline{x^{\prime\prime}}$ &$A_1+B_1$ & 4.48 & -- & 1.88, 3.77, 5.32, 6.33 & -- &  -- & --\\\hline
      Infrared & $E^\omega\parallel [001]$ & $A_2$ & -- & 2.58 & -- & -- &  -- & --\\
        & $E^\omega\parallel [1\bar{1}0]$  & $E_{\rm TO}$ / $E_{\rm LO}$ & -- & -- & -- & -- &  3.66, 5.37, 6.48 & 3.71, 5.90, 7.09\footnote{$E_{\rm LO}$ was determined by the peak position of the loss-function spectrum shown in Fig. 8(a).}\\\hline
      Terahertz\\Radiation & $E^\omega\parallel [010]$ & Unknown & -- & -- & -- & -- &  -- & 3.71\\
        \hline
\end{tabular}
\end{table*}

\begin{table}[t] 
\caption{Obtained fitting parameters of $\alpha$-TeO$_2$ in $E^\omega\parallel [001]$ configuration with $\epsilon_\infty=5.38$ and $E^\omega\parallel [1\bar{1}0]$ configuration with $\epsilon_\infty=5.25$ in the terahertz frequency region.}
    \begin{tabular}{cccccc}
      Representation &Mode $i$ & Configuration &$\omega_i$ (THz) & $\gamma_i$ (THz) & $f_i$ \\ \hline
      \hline
      $A_2$&1 & $E^\omega\parallel [001]$ &2.534 & 0.150 & 9.436 \\
      &2 &   &7.804 & 0.375 & 0.659 \\
      &3 &   &9.652 & 0.306 & 3.760\\
      &4 &   &18.14 & 0.685 & 2.873 \\\hline
      $E$ &5 & $E^\omega\parallel [1\bar{1}0]$ &3.660 & 0.111 & 0.654 \\
      &6 &   &5.314 & 0.155 & 6.958 \\
      &7 &   &6.400 & 0.361 & 2.404\\
      &8 &   &9.052 & 0.310 & 4.181\\
      &9 &   &9.797 & 0.495 & 0.574\\
      &10 &   &11.73 & 1.064 & 0.071\\
      &11 &   &18.83 & 0.561 & 1.655\\
      &12 &   &22.93 & 0.498 & 0.184\\\hline
\end{tabular}
\end{table}


\begin{thebibliography}{h}

\bibitem{MCardona} M. Cardona, in {\it Light Scattering in Solids: Basic Concepts and Instrumentation}, Springer Topics in Applied Physics {\bf 50} (Springer, 1982), pp. 19--178.


\bibitem{TDekorsyREV} T. Dekorsy, G. C. Cho, and H. Kurz, Coherent phonons in condensed media, in {\it Light Scattering in Solids VIII}, Springer Topics in Applied Physics {\bf 76}, pp. 169--209 (Springer, 2000).

\bibitem{RMerlin} R. Merlin, Solid State Commun. {\bf 102}, 207 (1997).

\bibitem{TDekorsy3} T. Dekorsy, W. Kutt, T. Pfeifer, and H. Kurz, Europhys. Lett. {\bf 23}, 223 (1993).

\bibitem{PCMPlanken} P. C. M. Planken, I. Brener, M. C. Nuss, M. S. C. Luo, and S. L. Chuang, Phys. Rev. B {\bf 48}, 4903 (1993).

\bibitem{AGambetta} A. Gambetta, C. Manzoni, E. Menna, M. Meneghetti, G. Cerullo, G. Lanzani, S. Tretiak,
A. Piryatinski, A. Saxena, R. L. Martin, and A. R. Bishop, Nat. Phys. {\bf 2}, 515 (2006).

\bibitem{MHase} M. Hase, M. Katsuragawa, A. M. Constantinescu, and H Petek, Nat. Photonics {\bf 6}, 243 (2012).


\bibitem{TDekorsy1} T. Dekorsy, H. Auer, C. Waschke, H. J. Bakker, H. G. Roskos, H. Kurz, V. Wagner, and P. Grosse, Phys. Rev. Lett. {\bf 74}, 738 (1995).


\bibitem{TDekorsy2} T. Dekorsy, H. Auer, C. Waschke, H. J. Bakker, H. G. Roskos, and H. Kurz, Phys. Rev. B {\bf 53}, 4005 (1996).


\bibitem{AVKuznetsov} A. V. Kuznetsov and C. J. Stanton, Phys. Rev. B {\bf 51}, 7555 (1995).

\bibitem{MNakayama} M. Nakayama, S. Ito, K. Mizoguchi, S. Saito, and K. Sakai, Appl. Phys. Express {\bf 1}, 012004 (2008).

\bibitem{MTani} M. Tani, R. Fukasawa, H. Abe, S. Matsuura, K. Sakai, and S. Nakashima, J. Appl. Phys. {\bf 83}, 2473 (1998).

\bibitem{PGu} P. Gu, M. Tani, K. Sakai, and T.-R. Yang, Appl. Phys. Lett. {\bf 77}, 1798 (2000).

\bibitem{PGu2} P. Gu, M. Tani, S. Kono, K. Sakai, and X.-C. Zhang, J. Appl. Phys. {\bf 91}, 5533 (2002).

\bibitem{MPHasselbeck} M. P. Hasselbeck, L. A. Schile, and D. Stalnaker, Appl. Phys. Lett. {\bf 85}, 173 (2004).

\bibitem{KTakeya} K. Takeya, Y. Takemoto, I. Kawayama, H. Murakami, T. Matsukawa, M. Yoshimura, Y. Mori, and M. Tonouchi, Europhys. Lett. {\bf 91}, 20004 (2010).


\bibitem{YXYan} Y-X. Yan, E. B. Gamble, and K. A. Nelson, J. Chem. Phys. {\bf 83}, 5391 (1985).

\bibitem{JLiebertz} J. Liebertz, Kristall. Technik {\bf 4}, 221 (1969).

\bibitem{NUchida} N. Uchida, Phys. Rev. B {\bf 4}, 3736 (1971).

\bibitem{PAThomas} P. A. Thomas, J. Phys. C: Solid State Phys. {\bf 21}, 4611 (1988).


\bibitem{RWGWyckoff} R. W. G. Wyckoff, {\it Crystal Structures 1} (Interscience Publishers, New York, 1963), pp239-444.

\bibitem{GArlt} G. Arlt and H. Schweppe, Solid State Commun. {\bf 6}, 783 (1968).

\bibitem{UchidaOhmachi} N. Uchida and Y. Ohmachi, J. Appl. Phys. {\bf 40}, 4692 (1969).

\bibitem{DSChemla} D. S. Chemla and J. Jerphagnon, Appl. Phys. Lett. {\bf 20}, 222 (1972).

\bibitem{SSingh} S. Singh, W. A. Bonner, and L. G. van Ultert, Phys. Lett. {\bf 38A}, 407 (1972). 

\bibitem{MTonouchi_Rev} M. Tonouchi, Nat. Photonics {\bf 1}, 97 (2007).

\bibitem{JFNye} J. F. Nye, {\it Physical Properties of Crystals} (Oxford University Press, England, 1976), p. 124.

\bibitem{YRShen} Y. R. Shen, {\it The Principles of Nonlinear Optics} (Wiley, New York, 1984).

\bibitem{TTakizawa} T. Takizawa, J. Phys. Soc. Jpn. {\bf 48}, 505 (1980).

\bibitem{JRobertson} J. Robertson, J. Phys. C: Solid State Phys. {\bf 12}, 4767 (1979).

\bibitem{DMKorn} D. M. Korn, A. S. Pine, G. Dresselhaus, and T. B. Reed, Phys. Rev. B {\bf 8}, 768 (1973).

\bibitem{ASPine} A. S. Pine and G. Dresselhaus, Phys. Rev. B {\bf 5}, 4087 (1972).

\bibitem{MKrauzman} M. Krauzman and J-P. Mathieu, Compt. Rend. Acad. Sci. Pari {\bf 273B}, 342 (1971).

\bibitem{BAyrault} B. Ayrault, E-A. Decamps, F. Abba, Y. Marqueston, and M. Durand, Solid State Commun. {\bf 11}, 639 (1972).

\bibitem{VRodriguez} V. Rodriguez, M. Couzi, F. Adamietz, M. Dussauze, G. Guery, T. Cardinal, P. Veber, K. Richardson, and P. Thomas, J. Raman Spectroscopy {\bf 44}, 739 (2013).

\bibitem{RKChang} R. K. Chang, J. Ducuing, and N. Bloembergen, Phys. Rev. Lett. {\bf 15}, 415 (1965).

\bibitem{HPWagner} H. P. Wagner, M. K\"{u}hnelt, W. Langbein, and J. M. Hvam, Phys. Rev. B {\bf 58}, 10494 (1998).

\bibitem{ANahata} A. Nahata, A. S. Weling, and T. F. Heinz, Appl, Phys. Lett. {\bf 69}, 2321 (1996).

\bibitem{ASchneider} A. Schneider, M. Neis, M. Stillhart, B. Ruiz, R. U. A. Khan, and P. G${\rm \ddot{u}}$nter, J. Opt. Soc. Am. B {\bf 23}, 1822 (2006).

\bibitem{CMTu} C. M. Tu, S. A. Ku, W. C. Chu, C. W. Luo, J. C. Chen, and C. C. Chi, J. Appl. Phys. {\bf 112}, 093110 (2012).

\bibitem{RTakeda} R. Takeda, N. Kida, M. Sotome, Y. Matsui, and H. Okamoto, Phys. Rev. A {\bf 89}, 033832 (2014).

\bibitem{TPDougherty} T. P. Dougherty, G. P. Wiederrecht, and K. A. Nelson, J. Opt. Soc. Am. B {\bf 9}, 2179 (1992).


\bibitem{YLiu} Y. Liu, A. Frenkel, G. A. Garrett, J. F. Whitaker, S. Fahy, C. Uher, and R. Merlin, Phys. Rev. Lett. {\bf 75}, 334 (1995).

\bibitem{KJYee} K. J. Yee, Y. S. Lim, T. Dekorsy, and D. S. Kimu, Phys. Rev. Lett. {\bf 86}, 1630 (2001).


\bibitem{SGPark} S-G. Park, M. R. Melloch, and A. M. Weiner, Appl. Phys. Lett. {\bf 73}, 3184 (1998).

\bibitem{SKono} S. Kono, M. Tani, and K. Sakai, Appl. Phys. Lett. {\bf 79}, 898 (2001).


\bibitem{SMatsuura} S. Matsuura, M. Tani, and K. Sakai, Appl, Phys. Lett. {\bf 70}, 559 (1997).


\bibitem{MTaniLT} M. Tani, K. Sakai, H. Abe, S. Nakashima, H. Harima, M. Hangyo, Y. Tokuda, K. Kanamoto, Y. Abe, and N. Tsukada, Jpn. J. Appl. Phys. {\bf 33}, 4807 (1994).


\bibitem{GGallot} G. Gallot, J. Zhang, R. W. McGowan, Tae-In Jeon, and D. Grischkowsky, Appl. Phys. Lett. {\bf 74}, 3450 (1999).














\end{thebibliography}
\end{document}